\documentclass[twocolumn,secnumarabic,amssymb, nobibnotes, aps, prd]{revtex4-1}

\newcommand{\macro}[1]{\texttt{\textbackslash#1}}
\newcommand{\m}[1]{\macro{#1}}

\usepackage[utf8]{inputenc}
\usepackage[T1]{fontenc}
\usepackage{lmodern}
\usepackage[english]{babel}
\usepackage{amsmath}
\usepackage{amssymb}
\usepackage{graphicx}
 \usepackage{epstopdf}
 \usepackage{listings}
 \usepackage{xcolor}
 \usepackage{subcaption}
\usepackage{float}
\usepackage{ulem}

\begin{document}
\title{Theoretical foundation of detrending methods for fluctuation analysis such as detrended fluctuation analysis and detrending moving average}%

\author{Marc H\"oll$^1$, Ken Kiyono$^2$, Holger Kantz$^3$}%
\affiliation{$^1$Department of Physics, Institute of Nanotechnology and Advanced Materials, Bar-Ilan University, Ramat-Gan, 52900, Israel\\
$^2$Graduate School of Engineering Science, Osaka University, 1-3 Machikaneyama-cho, Toyonaka, Osaka 560-8531, Japan\\
$^3$Max Planck Institute for the Physics of Complex Physics, N\"othnitzer Str. 38, 01187 Dresden, Germany}

\begin{abstract}

We present a general framework of detrending methods of fluctuation analysis
of which detrended fluctuation analysis (DFA) is one prominent
example. Another more recently introduced method is detrending moving average
(DMA). Both methods are constructed differently but are similarly able to
detect long-range correlations as well as anomalous diffusion even in the
presence of nonstationarities. In this article we describe their similarities
in a general framework of detrending methods. We establish this framework
independently of the definition of DFA or DMA but by investigating the failure
of standard statistical tools applied on nonstationary time series, let these
be intrinsic nonstationarities such as for Brownian pathes, or external ones due
to additive trends. In particular, we investigate the sample averaged mean squared
displacement of the summed time series. By modifying this estimator we
introduce a general form of the so-called fluctuation function and can
formulate the framework of detrending methods. A detrending method provides an
estimator of the fluctuation function which obeys the following principles: The first
relates the scaling behaviour of the fluctuation function to 
the stochastic properties of the time series. The second
principles claims unbiasedness of the estimatior. This is the centerpiece of
the detrending procedure and ensures that the detrending method can be applied
to nonstationary time series, e.g. FBM or additive trends. Both principles are
formulated and investigated in detail for DFA and DMA by
using the relationship between the fluctuation function and the
autocovariance function of the underlying stochastic process of the time
series.

\end{abstract}
\maketitle
\section{Introduction}\label{sec:introduction}
Long-range temporal correlations are omnipresent in a tremendous amount of
data sets and pose an ongoing challenge in time series analysis since its
first empirical detection by H.\ E.\ Hurst analyzing river flows
\cite{hurst}. For an excellent overview about the research history
see \cite{graves}. 
 Long-range correlations reflect some memory in a time series
recorded from a complex system and express themselves usually by a power-law
decay of the autocorrelation function. Especially in
nonstationary time series the existence of long-range correlations have a
non-negligible impact on the analysis, modelling and prediction of these
series, see \cite{palma,beran,pipiras}. Several traditional methods for the
detection of the correlation structure fail in the presence of
nonstationarities such as additive polynomial trends. These methods include
the sample estimation of the autocorrelation function, the R/S analysis
\cite{hurst} and the fluctuation analysis \cite{kantelhardt2}. Hence advanced
methods are required in order to detect reliably long-range correlations in
nonstationary time series. \\ \\
A very popular and frequently used method for the quantitative
characterizaton of long range correlations is detrended fluctuation
analysis (DFA) introduced by Peng et al.\ analysing DNA sequences \cite{peng}.
For a good introduction see \cite{kantelhardt2}. DFA has been applied to many
diverse fields such as heart rate variability
\cite{penzel,echeverria,castiglioni,baumert}, air temperature
\cite{talkner,kiraly,kurnaz,bunde,meyer,massah}, hydrology
\cite{hurst,zhang,zhang3}, cloud breaking \cite{ivanova}, sea surface temperature
\cite{luo}, stock prices \cite{cao,reboredo,serletis} and oil markets
\cite{ramirez,wang}.  The continuing success of DFA can be explained by its
easy construction as well as its well-perfoming results
\cite{taqqu2,hu,chen,chen2,xu,bashan,ma,weron}. In addition, DFA has been
developed further to also analyse multifractality
\cite{kantelhardt,movahed,movahed2,zhang2,zhou,zhou2} and cross-correlations
\cite{podobnik,zhou3,horvatic}. Another more recently developed method is
called detrending moving average (DMA), see \cite{alessio,arianos2} and for a good
overview see \cite{carbone4}. This method is also well-performing
\cite{xu,bashan,carbone2,carbone3,carbone5,shao} with strong applications in
analyzing financial data \cite{carbone6,dimatteo,serletis2,matsuhita,serletis3}
and fractal structures \cite{carbone7,carbone8,turk}. In \cite{kiyono4} a fast
algorithm has been proposed which drastically decreases the computation time
of DMA. Although DFA and DMA are constructed differently their basic
principles are working similarly in the time domain. We refer to these two
methods as detrending methods for fluctuation analysis in this article. 
Yet another powerful method is based on a wavelet-transform of a given time
series and was introduced by \cite{abry,veitch}, and we will later also
discuss its relationship to DFA and DMA. 
This relationship has already been investigated by \cite{kiyono5} for stationary processes.\\ \\
We focus here on detrending methods for fluctuation analysis which are modified random walk analysis
where the time series is interpreted as increment process of a random walk
like path. These methods provide an estimator of the so-called fluctuation
function whose scaling behaviour is directly related to the correlation type
of the time series. The implementation of detrending methods consists of
several straightforward steps transforming resulting in an estimator
of the fluctuation function. Hereby one crucial part of the procedure is the
trend elimination of the path in segments of the time axis, called
``detrending''. Since the detrending in DFA and DMA is ad hoc, it is by far not obvious how the
fluctuation function is connected to the correlation structure. \\ \\
The current analytical understanding of DFA and DMA is at a different states.
To our best knowledge analytical studies of DMA exist for the derivation of
the scaling behaviour for fractional Gaussian noise
\cite{arianos2,arianos,carbone} and on the ability of removing additive trends
\cite{carbone}. In contrast there exist relatively more analytical studies of
DFA which can be classified into four categories: 1) Calculation of the
scaling behaviour of the fluctuation function for specific process, namely
autoregressive model of first order \cite{hoell}, fractional Gaussian noise
\cite{hoell2,taqqu2,bardet,movahed2,crato} and FBM
\cite{movahed,heneghan,kiyono}; 2) Derivation of the relationship between the
fluctuation function and known statistical quantities, namely the
autocorrelation function \cite{hoell2}, power spectrum
\cite{heneghan,kiyono,kiyono2,kiyono3,talkner,willson,willson2} and variogram
\cite{lovstetten}; 3) Describing statistical properties of the fluctuation
function \cite{bardet,crato}; 4) Illuminating the functionality of detrending
\cite{hoell3}. Nevertheless there are still many open questions about these
methods, see for example \cite{kiyono3,bryce}, not just about
minor technical details but questions about fundamental principles and
properties of detrending methods. We know that detrending methods work for a
large class of nonstationary processes but we are ignorant of why they
work. This is actually rather suprising since the operation of detrending is
the centrepiece of detrending methods. Since detrending is usually done on
segments of the time series, it leads to mutually inconsistent local trends,
and never reproduces a given global, e.g., linear trend on data.\\ \\
In this article, we present an intuitive and natural motivation of detrending
methods and also demonstrate in detail how they work for different types of
nonstationarities. In order to accomplish this goal it is essential to derive
the general relationship between the fluctuation function and the
autocovariance function. Hence this article is constructed as follows. In
Sec.\ 2 we recapitulate problems of the sample autocovariance function
in the presence of long-range correlations and nonstationarities. We argue
that the mean squared displacement of the random walk path provides a better
tool to analyze the correlation structure of a time series.  In Sec.\ 3 we
show that the estimator of this mean squared displacement requires stationary
increments which is only fulfilled for stationary time series. We demonstrate
the failure of the estimator in analyzing the scaling behaviour for two
different types of nonstationarity, namely intrinsic and external
nonstationarity. In Sec.\ 4 we introduce the general framework of detrending
methods and formulate two basic principles which must be fulfilled in order to
estimate reliably the correlation structure. In Sec.\ 5 we discuss that 
 DFA and DMA are indeed specific realizations of the general framework of 
detrending methods. Here we only
focus on centered DMA and not backward and forward DMA. In Sec.\ 6 we explicitely show that DFA and DMA fulfill
the two principles of detrending methods. 
\section{Basics and motivation}
\subsection{Autocorrelation function}\label{sec:acfintro}
Given is a time series $\{ x(t)\}_{t=1}^N$. We understand this time series as
a superposition of a single realisation of a stochastic process
$\{\epsilon(t)\}_{t=1}^N$ and a deterministic function $\{m(t)\}_{t=1}^N$
given as  
\begin{equation}\label{timeseries1}
x(t) = \epsilon(t) + m(t).
\end{equation}
We restrict ourselves to Gaussian stochastic processes with time independent
zero mean. 
$\{\epsilon(t)\}_{t=1}^N$ is a realisation of an either discrete or a
continuous process, where the latter is sampled at discrete
times.  The stochastic process itself can be stationary or
nonstationary. For the sake of simplicity we use in the following the notation
$\{ \epsilon(t)\}_{t=1}^N$ for either a stochastic process or a single
realisation and mention accordingly which of them we mean. Therefore we also
use $\{ x(t)\}_{t=1}^N$ for the combination of a deterministic function $m(t)$
and either a stochastic process or a single realisation. Usually, $m(t)$ is a
polynomial in $t$ 
but it can also be any other non-stochastic
function such as being periodic. \\ \\

Let us first assume that the stochastic process is stationary. The autocorrelation function 
\begin{equation}\label{defcorr}
C(\tau)  = \frac{\text{Cov}(\tau)}{\text{Cov}(0)} = \frac{\langle \epsilon(t) \epsilon(t+\tau) \rangle}{\langle \epsilon^2(t) \rangle}
\end{equation}
is a relevant characteristic. If the stochastic
process is nonstationary then the autocovariance function depends on both time
points 
$\text{Cov}(i,j)=\langle \epsilon(i)\epsilon(j)\rangle$. 

There are two important classes of correlation types depending on the
behaviour of the autocorrelation function for large time lags $\tau$:
short-range correlations and long-range correlations. A short-range correlated
processes has a finite characteristic correlation time $s_c=\int_0^\infty
C(\tau)\mathrm{d}\tau<\infty$ which is characterised by the convergence of the
sum of the autocorrelation function over all time lags $\sum_{\tau=0}^\infty
C(\tau) < \infty$. Notably the uncorrelated white noise process with zero mean
and unit variance is included in the class of short-range correlated
processes. If on the other the sum diverges $\sum_{\tau=0}^\infty C(\tau) =
\infty$ then the process is long-range correlated. Hence there exists no
characteristic correlation time. Such processes forget their initial
conditions very slowly which is known as long memory. Long-range correlations
are often described by a decreasing power law of  the autocorrelation function

\begin{equation}\label{lrc}
C(\tau) \sim \tau^{-\gamma}
\end{equation}
with correlation parameter $0<\gamma<1$. An important theoretical model is
fractional Gaussian Noise (FGN), see \cite{mandelbrot}. FGN is the stationary
increment process of the self-similar fractional Brownian motion (FBM) with
self-similarity parameter $H$. This parameter is often called Hurst parameter
and can have the values $1/2 < H < 1$. We exclude here the anticorrelated
regime $0<H<1/2$. Note that the self-similarity parameter and the Hurst
parameter are actually different \cite{chen3} but for the here studied
processes with Gaussian and stationary increments both are equivalent. The
relationship between $H$ and the correlation parameter of Eq. (\ref{lrc}) is
$\gamma=2-2H$. In the special case of $H=1/2$ FBM is the standard Brownian
motion (BM).\\ \\

In order to decide weither or not a given time series is long-range correlated
the autocorrelation function in Eq.(\ref{defcorr}) has to be estimated. This
can be done straightforwardly with the sample estimator of the autocorrelation
function. The numerator and denominator in Eq.(\ref{defcorr}) is estimated
with the sample estimator of the autocovariance function 
\begin{equation}\label{standardest}
\widehat{\text{Cov}}_\text{sample} (\tau) = \frac{1}{N-\tau} \sum_{i=1}^{N-\tau}x(i)x(i+\tau)
\end{equation}
where products of the time series $x(i)x(i+\tau)$ are averaged over all
possible time points $i$ for a given time lag $\tau$. Usually one replaces
$x(t)$ by $x(t)-\bar{x}(t)$ with the sample mean $\bar{x}(t)$. Unfortunately
the estimator $\widehat{\text{Cov}}_\text{sample} (\tau)$ has at least two
important estimation problems: 
\begin{itemize}
\item[(E1)] The estimator $\widehat{\text{Cov}}_\text{sample} (\tau)$ fluctuates strongly around zero for large $\tau$. This is even the case for only positive values of the true autocovariance function. This statistical uncertainty makes it very difficult to observe a power law $\tau^{-\gamma}$ in the log-log plot which is even worse for short time series.
\item[(E2)] The estimator $\widehat{\text{Cov}}_\text{sample} (\tau)$  is only
  meaningful for stationary time series. It can be furthermore even misleading
  in the sense that this estimator applied to the supperposition of a
  stationary short-range correlated stochastic processes and a linear trend
  can show a power law behaviour. This will eventually be misinterpreted as a
  stationary long-range correlated process, see \cite{hoell,maraun}. 
\end{itemize}
A direct estimation of the autocorrelation function is often not possible due
to the estimation problems (E1) and (E2). Hence it is reasonable to gain
indirectly access to the correlation behaviour using different approaches. We
introduce in this article our framework of so-called detrending methods as
possible solution of the estimation problems (E1) and (E2). DFA and DMA are
examples of these detrending methods. Although it is known that these methods
can overcome the problems (E1) and (E2) it lacks until today of both an
intuitive understanding as well as a rigorous description. We present in the
following basic ideas deduced from well-known mathematical functions and hope
we can contribute to the fundamental understanding of the nature of detrending
methods.

\subsection{Mean squared displacement of the path}\label{sec:msd}

\begin{figure}
  \includegraphics[width=1.05\linewidth]{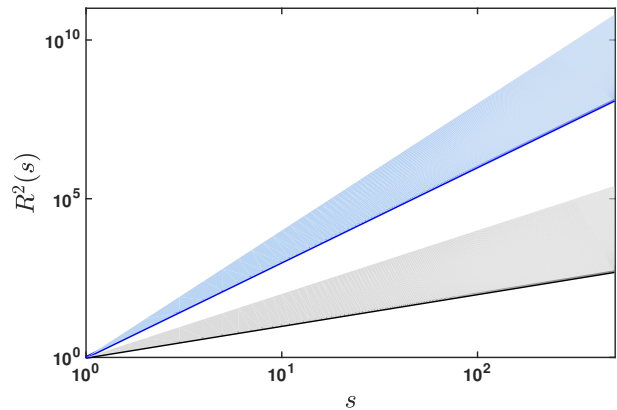}
  \caption{\small{Scaling behaviour of the path MSD for stochastic processes discussed in this article. For the sake of simplicity we used a prefactor of $1$ in Eq.~(\ref{scalmsd1}) for this figure, i.e. $R^2(s)=s^{2\alpha}$. The figure shows white noise with $\alpha=1/2$ (black line), fractional Gaussian noise with $1/2<\alpha<1$ (grey shaded area), Brownian motion with $\alpha=3/2$ (blue line) and fractional Brownian motion with $3/2<\alpha<1$ (blue shaded area). Note that we only consider specific values of the Hurst parameter $H$ here, see Eq.~(\ref{disting}) for a detailed classification of the here discussed processes.}}
  \label{fig:01}
\end{figure}

To tackle the first estimation problem (E1) we consider the mean squared
displacement (MSD) of the path of the stochastic process $\{r(t)\}_{t=1}^N$. 
We define this path as the cumulative sum of the stochastic process
\begin{equation}\label{profiledef}
r(t) = \sum_{i=1}^t \epsilon(i)
\end{equation}
so that the stochastic process $\{\epsilon(t)\}_{t=1}^N$ is the increment
process of $\{r(t)\}_{t=1}^N$. If the stochastic process is WN then the path
is the standard random walk. The correlation type of the stochastic process,
assuming a finite second moment $\langle \epsilon^2(t)\rangle$ and
stationarity, 
can be directly connected to the scaling behaviour in $s$ of the path MSD
\begin{equation}\label{msddef}
R^2(s)=\langle (r(s)-r(0))^2 \rangle = \langle r^2(s) \rangle
\end{equation}
which is the mean of the squared displacement with $r(0)=0$. We call this
equation the path representation of the path MSD because we will later also
introduce a second representation with respect to the increments
$\{\epsilon(t)\}_{t=1}^N$. Here $s$ is the time covered by the path. In
numerical estimates, it 
should be less than the length of the time series $s\le N$. The path MSD
scales increasingly like 
\begin{equation}\label{scalmsd1}
R^2(s) \sim s^{2\alpha}
\end{equation}
with the so-called fluctuation parameter $\alpha$ which will be explained as
follows. First we consider stationary stochastic processes. If the stochastic
processes is white noise (WN) then the path is BM. Hence the path MSD scales
linearly, $R^2(s) \sim s$. If the time series is FGN with $1/2<H<1$ then the
path is FBM. Hence the path MSD scales super-diffusively $R^2(s) \sim
s^{2H}$. Although we present here only WN and FGN, processes with an
autoregressive part have the same asymptotic scaling behaviour. They can be
modeled by an AR($1$) process in the case of short-range correlations and an
ARFIMA($1$,$d$,$0$) process in the case of long-range correlations. But then
only for large enough $s$ the scaling of $\alpha$ is described as above due to
the existence of a larger scaling regime for small $s$. This crossover
behaviour requires large data sets in order to observe the correct
scaling. Summarizing, for stationary stochastic processes the knowledge of
$\alpha$ in Eq.~(\ref{scalmsd1}) allows us to distinguish between short-range and
long-range correlations.\\ \\ 
Interestingly, the path MSD contains also information when the stochastic
process is nonstationary, whereas for the autocorrelation function stationarity
has to be assumed. This means that the path MSD can show anomalous diffusive
scaling. We depict this with BM and FBM. If the series is BM then the path is
summed BM. Hence the path MSD scales cubically $R^2(s)\sim s^3$, see
\cite{hoell}. If the series is FBM with $1/2<H<1$ then the path is summed
FBM. Hence the path MSD scales like $R^2(s)\sim s^{2H+2}$, see
\cite{taqqu2}. We exclude here the anticorrelated case $0<H<1/2$. So the
scaling of the path MSD $R^2(s) $ allows us to distinguish between a large
class of stochastic processes, stationary and nonstationary, by
the value of the fluctuation parameter 
\begin{equation}\label{disting}
\alpha \in
\begin{cases}
\{ 1/2 \} & \text{ for WN}, \\
(1/2,1) & \text{ for FGN with } \alpha=H,\\
\{ 3/2 \} & \text{ for BM},\\
(3/2,2) & \text{ for FBM with } \alpha=H+1
\end{cases}
\end{equation}
with $1/2<H<1$, see figure \ref{fig:01}. To be precise, the type of
nonstationarity which is exhibited by BM and FBM is called intrinsic, see
\cite{hoell3}. This means that for a pure noise driven time series
$x(t)=\epsilon(t)$ the nonstationarity comes from the nonstationary stochastic
process $\{\epsilon(t)\}_{t=1}^N$. A second type of nonstationarity introduced
in \cite{hoell3} is the external one. This means that for the composed time
series $x(t)=\epsilon(t)+m(t)$ the nonstationarity comes from the
deterministic function $\{m(t)\}_{t=1}^N$ when the stochastic process
$\{\epsilon(t)\}_{t=1}^N$ is stationary. Let us consider the example of a time
series composed of WN or FGN superimposed by a linear trend $m(t) \propto t$. The
path MSD here still scales as in Eq.~(\ref{disting}) because the path is
defined as the sum of the stationary process
$r(t)=\sum_{i=1}^t\epsilon(i)$. But would the path be defined as sum of the
time series instead of the stochastic process, nameley as
$\tilde{r}(t)=\sum_{i=1}^tx(i)$, then the path MSD would scale ballistically $R^2(s)
\sim s^4$ so that $\alpha=2$. In any applications, one is interested in
estimating properties of the stochastic process even in the presence of
additive deterministic functions. Hence we define the path as sum of
the stochastic process and not the full time series. An appropriate estimator
of the path MSD should therefore also scale as in Eq.~(\ref{disting}) and
ignore the presence of determistic additive signal components. The naive
estimators instead, calculating the MSD of  $\tilde{r}(t)=\sum_{i=1}^tx(i)$,
in such cases would show ballistic scaling and the information about
the stochastic process would be lost.

\subsubsection{Increment represenation of the path MSD}
In Eq.~(\ref{msddef}) the path MSD $R^2$ is defined as the squared displacement of the path. We also can express $R^2$ in relationhip with the autocovariance function
\begin{equation}\label{fullmsd}
R^2(s) = \sum_{i,j=1}^s \text{Cov}(i,j)
\end{equation}
which is derived by taking the square of the right hand side of Eq.~(\ref{profiledef}). We call this equation the increment representation of the path MSD whereas its definition in Eq.~(\ref{msddef}) is called path representation. The increment representation is useful to understand the above discussed scaling of $R^2(s)$. For nonstationary FBM we refer to \cite{taqqu2} where the relationship $\alpha=H+1$ has been straigthforwardly calculated with the help of the right hand side of Eq.~(\ref{fullmsd}). For stationary processes it is possible to give a more intuitive understanding of the scaling behaviour of the path MSD and in particular why the path MSD or better an unbiased estimator of the path MSD is able to overcome the estimation problem (E1). First we order Eq.~({\ref{fullmsd}) according to the time lag $\tau$
\begin{equation}\label{orderedtimelag}
R^2(s) = \sum_{i=1}^s \text{Cov}(0) + 2\sum_{\tau=1}^{s-1} \sum_{i=1}^{s-\tau}\text{Cov}(\tau).
\end{equation}
Since the autocovariance function $\langle \epsilon(i) \epsilon(i+\tau) \rangle$ doesn't depend on the time point $i$ we can put the sum over $i$ to the right
\begin{equation}
\label{kernelFA}
R^2(s) =\text{Cov}(0)\left(s + 2\sum_{\tau=1}^{s-1}C(\tau)(s-\tau) \right).
\end{equation}
Inserting specific autocorrelation functions in this equation verifies the above explained scaling behaviour of $R^2(s)$ for stationary stochastic processes, see \cite{kantelhardt2}. We leave the detailed derivation out here since we present later a more general equation and verify the scaling there. \\ \\
Finally with Eq.~(\ref{kernelFA}) we can understand why the path MSD overcomes
the estimation problem (E1), see \cite{kantelhardt2}: The sum
$\sum_{\tau=1}^{s-1}C(\tau)(s-\tau) $ in Eq.~({\ref{kernelFA}) scales in the
  same way as
  the integral $\int_{0}^sC(\tau)(s-\tau)\text{d}\tau$ which represents the
  area under $C(\tau)(s-\tau)$. Whereas the autocorrelation function decays in
  $s$, the integral increases, specifically with a power
  law decay $C(s)\sim s^{-\gamma}$ the integral scales like $s^{2-\gamma}$. 
  Hence, the MSD overcomes (E1) because for
  large $s$ the values of also large, $s^{2-\gamma}$, and fluctuate only weakly
  in the log-log plot.

\subsubsection{Estimation of the path MSD}\label{sec:estimsd}
We are in principle able to overcome the estimation problem (E1), but we have
to find an estimator $\widehat{R^2}(s)$ of the path MSD $R^2(s)$ which can be
applied to a given time series $\{x(t)\}_{t=1}^N$. This estimator should also
be able to overcome estimation problem (E2). But this will be a difficult task
which leads to a modified version of $\widehat{R^2}(s)$ and the introduction
of detrending methods. \\ \\ 
The increment representation of Eq.~(\ref{fullmsd}) directly connects the path MSD and the autocovariance function, namely by $R^2(s)  = \sum_{i,j=1}^s \text{Cov}(i,j)$. We use this connection to connect similarly an estimator of the path MSD $\widehat{R^2}(s)$ to an estimator of the autocovariance function $\widehat{\text{Cov}}(i,j)$, namely by
\begin{equation}
 \widehat{R^2}(s)  = \sum_{i,j=1}^s \widehat{\text{Cov}}(i,j)
\end{equation}
which is the increment represenation of the estimator of the path MSD. Below
we will try to find a suitable estimator $\widehat{R^2}(s)$ in detail because
this investigation serves as background of the motivation and formulation of
detrending methods. \\ \\ 
Although the autocovariance function is a function of the stochastic process
$\{\epsilon(t)\}_{t=1}^N$, the estimator $\widehat{\text{Cov}}(i,j)$ is
applied to the time series $\{x(t)\}_{t=1}^N$ and not the single realisation
of the stochastic process. Possible
external influences $\{m(t)\}_{t=1}^N$ are unknown a priori and might be part of the
time series. Therefore we need to formulate an equivalent of the path for
further investigation. The path is defined as the cumulative sum
$r(t)=\sum_{i=1}^t \epsilon(i)$ of the stochastic process. Similarly we define
the path of the time series as cumulative sum of the time series 
\begin{equation}
y(t)=\sum_{i=1}^t x(i).
\end{equation}
We also call this the profile. So the time series is understood as increment process of the profile. In the case of a pure random time series the profile is the path $y(t)=r(t)$. Similar to the path MSD $R^2=\langle r^2(s)\rangle$ we have now the profile MSD as $\langle y^2(s)\rangle$ which are identical again only for pure random time series.
\section{Pointwise averaging procedure}
No matter if the time series is stationary or not, we simply use now the sample autocovariance function as estimator for the autocovariance function and study the consequences of our choice. The sample autocovariance function only depends on the time lag $\tau$ so we can use Eq.~(\ref{kernelFA}) and find $\widehat{R^2}(s)$ by replacing $\text{Cov}(\tau)$ with $ \widehat{\text{Cov}}_\text{sample}(\tau)$, it follows
\begin{equation}
\widehat{R^2}_\text{sample}(s) =   \widehat{\text{Cov}}_\text{sample}(0) s + 2 \sum_{\tau=1}^{s-1} \widehat{\text{Cov}}_\text{sample}(\tau) (s-\tau).
\end{equation}
The notation "sample" as index of $\widehat{R^2}_\text{sample}(s)$ indicates the averaging procedure used for the estimation of the autocovariance function. If we use now the definition of the sample autocovariance function $\widehat{\text{Cov}}_\text{sample}(\tau)=1/(s-\tau)\sum_{i=1}^{s-\tau} x(i) x(i+\tau)$ then we find
\begin{equation}
\widehat{R^2}_\text{sample}(s) = \sum_{i,j=1}^s x(i) x(j).
\end{equation}
but this is exactly the squared profile
\begin{equation}\label{fsdagf}
\widehat{R^2}_\text{sample}(s) =  y^2(s).
\end{equation}
This result has two problems. First it does not estimate the path MSD but the
profile MSD. Hence it will suffer from the unwanted influences of external nonstationarities.
But this is expected since we simply used the sample
autocovariance. Also the below introduced estimator will have problems with
nonstationarity. The second and more obstructive problem of Eq.~(\ref{fsdagf})
is that the single value $y^2(s)$ is not a reliable estimate. Therefore we
present in the next section a better estimation technique which is also used
by detrending methods. 

\section{Segmentwise averaging procedure}\label{sec:segwise}
\subsection{Segmentation of the time axis and estimator of the autocovariance function}\label{sec:segwise2}
In order to find a better estimator of the autocovariance function than in the
previous subsection we have to replace the ensemble average in a suitable
way. The meaning of the ensemble average of a random variable is that we
ideally average over an infinite amount of samples. Therefore the
autocovariance function can be understood as 
\begin{equation}\label{ensembledef}
\text{Cov}(i,j) = \langle \epsilon(i)\epsilon(j)\rangle=  \lim_{K \to \infty} \frac{1}{K} \sum_{\nu=1}^K \epsilon^{(\nu)}(i) \epsilon^{(\nu)}(j).
\end{equation}
Here $\{\epsilon^{(\nu)}(i)\}_{i\in\mathbb{N}}$ represents the $\nu$-th
realisation of the stochastic process $\{\epsilon(i)\}_{i\in\mathbb{N}}$. In 
practice, the limit $K\to\infty$ cannot be performed, 
since we usually only
have a finite sample, or as in this article, only one single realisation with finite
length of the time axis. Given only one realisation
$\{\epsilon^{(1)}(i)\}_{i=1}^N$ we replace the ensembles in
Eq.~(\ref{ensembledef}) appropriately using a segmentation of the time axis
which is done as follows.\\ \\ 
We divide the time axis $[1,N]$ into $K\in\mathbb{N}$ segments of length
$s$. The $\nu$-th segment consists of the time points $[1+d_v,s+d_v]$ with
$\nu\in\{1,K\}$. The quantity $d_v$ shifts time points from the first to the
$\nu$-th segment 
\begin{equation}
i \in [1,s] \Rightarrow i+ d_v \in [1+d_v,s+d_v]
\end{equation}
and we therefore call $d_\nu$ shift factor which also depends on $s$. For the
first segment $\nu=1$ we claim $d_1=0$. For the last segment $\nu=K$ the
inequality has to be hold $s+d_K \le N$. Note that such a segmentation of the
time axis $[1,N]$ usually has some leftovers which are not in any of the
segments $[1+d_\nu,s+d_\nu]$. For the sake of simplicity we don't treat them
here. Time points can be shifted from the first segment to the $\nu$-th one in
many possible ways depending on the exact form of the shift factor $d_\nu$. We
discuss now two important ones. First the segments can be shifted subsequently
so that we have the segments 
\begin{equation}
[1,s], [2,s+1], [3,s+2], \ldots, [1+d_K,s+d_K].
\end{equation}
Here the shift factor is $d_\nu=\nu-1$ and we have $K=N-s+1$ segments. Alternatively,
the segments can be shifted with distance $s$ such that we have the disjoint
segments 
\begin{equation}
[1,s], [s+1,2s], [2s+1,3s], \ldots, [1+d_K,s+d_K].
\end{equation}
Here the shift factor is $d_\nu=(\nu-1)s$ and we have $K=\lfloor N/s\rfloor$
segments with the floor function $\lfloor \ldots \rfloor$. Both shifting
procedures are later used in this article. \\ \\ 
This segmentation of the time axis helps us to estimate the autocovariance function of Eq.~(\ref{ensembledef}) when only one realisation $\{\epsilon^{(1)}(i)\}_{i=1}^N$ is given as it is often the case in time series analysis. We cut this realisation into segments and take these as ensemble \\ \\
\begin{equation}\label{replace}
\{\epsilon^{(1)}(i+d_\nu)\}_{i=1}^s \longleftrightarrow  \{\epsilon^{(\nu)}(i)\}_{i=1}^s,
\end{equation}
see figure \ref{fig:02}. For the sake of simplicity we skip the symbol the
first realisation $(1)$ in the following $\epsilon^{(1)}(i) \rightarrow
\epsilon(i)$ and hence go back to our previous notation. We can now define an
estimator of the autocorrelation function by using the replacement of
Eq.~(\ref{replace}) for the autocovariance function of Eq.~(\ref{ensembledef})
for finite $K$, namely 
\begin{equation}\label{segfinal0}
\widehat{\text{Cov}}_\text{seg}(i,j)= \frac{1}{K} \sum_{\nu=1}^K \epsilon(i+d_\nu)\epsilon(j+d_\nu).
\end{equation}
The index "seg" stands for segment. This estimator is only unbiased for
stationary stochastic processes
$\langle\widehat{\text{Cov}}_\text{seg}(i,j)\rangle=\text{Cov}(i,j)$ but not
for nonstationary processes. But this is not suprising because the replacement
of ensembles with the realisation in segments is only reasonable for
stationarity. Nevertheless the here introduced segmentwise estimator of the
autocovariance function will play an important role in detrending methods and
we will therefore continue to investigate it.\\ \\ 
Due to construction this estimator is applied to a pure random time series
$x(t)=\epsilon(t)$ which is of course unknown a priori. An actual practical
estimator is applied on the full time series and we therefore can finally
define the segmentwise averaging estimator of the autocovariance function as 
\begin{equation}\label{segfinal}
\widehat{\text{Cov}}_\text{seg}(i,j)= \frac{1}{K} \sum_{\nu=1}^K x(i+d_\nu)x(j+d_\nu).
\end{equation}
This estimator implies that the the $\nu$-th
segment of the single given time series 
represents the $\nu$-th realisation of an assumed existing ensemble of
time series. Using this result we will find next an estimator of the path
MSD.

\begin{figure}[H]
  \includegraphics[width=1\linewidth]{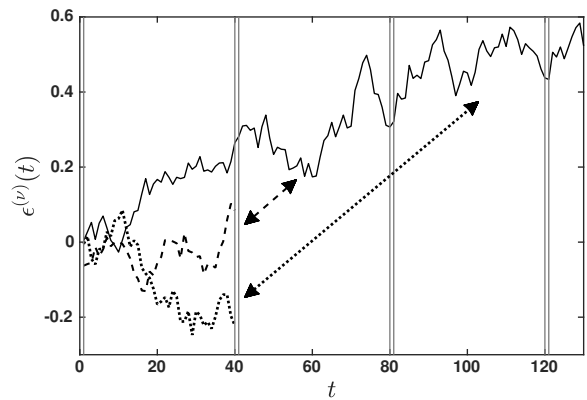}
\caption{\small{Schematic representation of the of the replacement of the
    $\nu$-th realisation in the first segment
    $\{\epsilon^{(\nu)}(i)\}_{i=1}^s$ by the first realisation in the $\nu$-th
    segment $\{\epsilon^{(1)}(i)\}_{i=1+d_\nu}^{s+d_\nu}$ with
    $\nu\in\{1,K\}$, see Eq.~(\ref{replace}). It is shown here for $K=3$
    realisations/segments with Brownian motion as stochastic process. The
    first realisation (solid line) represents the time series, the second
    (dashed line) and third realisation (dotted line) are assumed to exist
    theoretically in the first segment, see more details in the text of this
    section. Here the segments have a length of $s=40$ and are disjoint with
    shift factor $d_\nu=(\nu-1)s$.}}\label{fig:02} 
\end{figure}

\subsection{Estimator of the path MSD}
Now we take the relationship between the path MSD and the autocovariance
function $R^2(s)=\sum_{i,j=1}^s\text{Cov}(i,j)$ of Eq.~(\ref{fullmsd}) and
replace the autocovariance function with the segmentwise averaging estimator
of Eq.~(\ref{segfinal}). So we obtain the estimator of the path MSD as 
\begin{equation}\begin{split}\label{defestr2}
\widehat{R^2}_\text{seg}(s) &= \sum_{i,j=1}^s \widehat{\text{Cov}}_\text{seg}(i,j)\\
&= \frac{1}{K} \sum_{\nu=1}^K\sum_{i,j=1}^s x(i+d_\nu)x(j+d_\nu).
\end{split}\end{equation}
In the second step we put the sum over the segments to the left. The notation
"seg" as index of $\widehat{R^2}_\text{seg}(s)$ indicates the segmentwise
averaging procedure used for the estimation of the autocovariance
function. This equation is the increment representation of the estimator
$\widehat{R^2}_\text{seg}(s)$. The path representation can be derived with
respect to the profile of the time series. The double sum of the time series
over $i$ and $j$ in Eq.~(\ref{defestr2}) can be written as the squared sum of
the time series 
\begin{equation}
\sum_{i,j=1}^s x(i+d_\nu)x(j+d_\nu) = \left( \sum_{i=1}^s x(i+d_\nu) \right)^2.
\end{equation}
Inside this square the sum of the time series is equivalent to the profile displacement
\begin{equation}\label{deffidispli}
\sum_{i=1}^s x(i+d_\nu)=\sum_{i=1}^{s+d_\nu} x(i)-\sum_{i=1}^{d_\nu} x(i)=y(s+d_\nu)-y(d_\nu),
\end{equation}
see figure \ref{fig:03}. So the sum of the time series over all points in the
segment $\nu$ of length $s$ is the same as the displacement of the profile
where the profile starts at $t=d_\nu$ and walks a time interval of $s$ until
$t=s+d_\nu$. Using this connection we can write the estimator
$\widehat{R^2}_\text{seg}(s)$ in Eq.~(\ref{defestr2}) as average of the
squared displacements of the profile 
\begin{equation}\label{msdestiii}
\widehat{R^2}_\text{seg}(s) = \frac{1}{K} \sum_{\nu=1}^K (y(s+d_\nu)-y(d_\nu))^2
\end{equation}
where we average over the number of segments. We call this equation the path representation of the estimator. \\ \\
The estimator of the path MSD is a general form of two well-known quantities
depending on the shift factor $d_\nu$. For the subsequent shift factor
$d_\nu=\nu-1$ the estimator $\widehat{R^2}_\text{seg}(s)$ is the time averaged
mean squared displacement (TAMSD) of the profile 
\begin{equation}
\widehat{R^2}_\text{seg}(s) = \frac{1}{K} \sum_{\nu=0}^{K-1} (y(s+\nu)-y(\nu))^2
\end{equation}
with $K=N-s+1$ where we decremented the summation index $\nu$ by $1$. Note
that $y(0)=0$. A modification of this estimator will later lead to the method
of detrending moving average. In the other case described above, the disjoint
segmentation  with shift factor $d_\nu=(\nu-1)s$ the estimator
$\widehat{R^2}_\text{seg}(s)$ is the fluctuation function of the method of
fluctuation analysis 
\begin{equation}
\widehat{R^2}_\text{seg}(s) = \frac{1}{K} \sum_{\nu=0}^{K-1} (y((\nu+1) s)-y(\nu s))^2
\end{equation}
with $K=\lfloor N/s \rfloor$, see \cite{kantelhardt2}. Here we also
decremented the summation index $\nu$ by $1$. A modification of this
fluctuation function with the help of Eq.~(\ref{defestr2}) will later lead to
the method of detrended fluctuation analysis. \\ \\ 
Summarized, the here introduced estimator $\widehat{R^2}_\text{seg}(s)$ allows
later a simple modification which leads to the basic principle of detrending
of detrending methods. This modification is applied to
$\widehat{R^2}_\text{seg}(s)$ in the increment representation of
Eq.~(\ref{defestr2}). A modification with the help of the path representation
of Eq.~(\ref{msdestiii}) is not always possible, e.g., not to derive detrended fluctuation
analysis DFA.

\begin{figure}
  \includegraphics[width=1\linewidth]{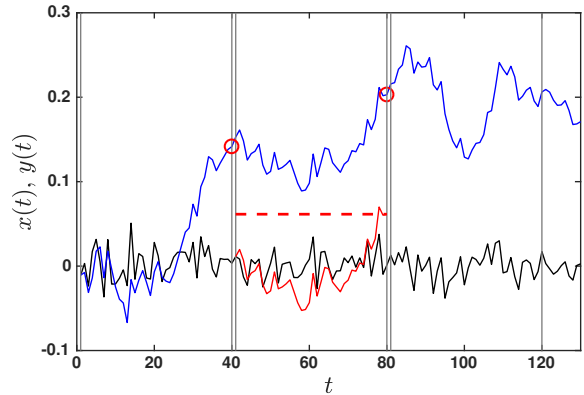}
  \caption{\small{Schematic representation of the equivalence between  the
      profile displacement $y(s+d_\nu)-y(d_\nu)$ and the cumulative sum of the
      time series $\sum_{i=1}^s x(i+d_\nu)$, see Eq.~(\ref{deffidispli}). We
      used white noise (black line)/Brownian motion (blue line) as time
      series/profile and disjoint segments of length of $s=40$ with the shift
      factor $d_\nu=(\nu-1)s$. The figure highlights the relationship in the
      second segment $\nu=2$ where we display the time series $\{\sum_{i=1}^t
      x(i+d_2)\}_{t=1}^s$ (red line) and the two points of the profile
      $y(d_2)$ and $y(s+d_2)$ (two red dots). The relationship of
      Eq.~(\ref{deffidispli}) says that  $\sum_{i=1}^s x(i+d_2)$ and
      $y(s+d_2)-y(d_2)$ are equivalent which is indicated with the red dashed
      line in this figure.}} 
  \label{fig:03}
\end{figure}

\subsection{Bias of the estimator of the path MSD}
The bias of the estimator $\widehat{R^2}_\text{seg}(s)$ is given by the
difference between the mean of the estimator of the path MSD and the path MSD 
\begin{equation}
B(s) = \langle \widehat{R^2}(s) \rangle - R^2(s).
\end{equation}
In order to derive the bias in the case of the segmentwise averaging procedure
we need to calculate the mean of the estimator $\widehat{R^2}_\text{seg}(s)$
for which we use the increment representation 
\begin{equation}\label{calku}
\langle\widehat{R^2}_\text{seg}(s)\rangle  =  \frac{1}{K} \sum_{\nu=1}^K \sum_{i,j=1}^s \langle x(i+\delta_\nu)x(j+\delta_\nu) \rangle.
\end{equation}
Below we will analyze $\langle\widehat{R^2}_\text{seg}(s)\rangle$ depending on
the stationarity of the time series. We will see that the estimator is only
unbiased for stationary processes but not for nonstationary ones. 

\subsection{Stationary time series}
Here we analyze the bias of the estimator $\widehat{R^2}_\text{seg}(s)$ for stationary time series. When the time series is stationary, then our model is a pure random time series $x(t)=\epsilon(t)$ because there can be no additive trends $m(t)=0$. \\ \\
Using the increment representation of Eq.~({\ref{calku}) we find for the mean of the estimator
\begin{equation}\label{meanfluc}
\langle\widehat{R^2}_\text{seg}(s)\rangle  = \sum_{i,j=1}^s \text{Cov}(i,j)
\end{equation}
which is exactly the path MSD $R^2(s)$. We used that the stationary autocovariance function is independent of the segment $\langle \epsilon(i+\delta_\nu)\epsilon(j+\delta_\nu) \rangle=\langle \epsilon(i)\epsilon(j) \rangle$. So the estimator is unbiased
\begin{equation}
\langle\widehat{R^2}_\text{seg}(s)\rangle = R^2(s).
\end{equation}
The same result can be obtained with the path representation
\begin{equation}
\langle\widehat{R^2}_\text{seg}(s) \rangle = \frac{1}{K} \sum_{\nu=1}^K \langle(r(s+d_\nu)-r(d_\nu))^2\rangle= \langle r^2(s)\rangle
\end{equation}
which is also the path MSD. We used that the displacements of the path are stationary. \\ \\
In summary, the estimator $\widehat{R^2}_\text{seg}(s)$ estimates the path MSD in the case of a pure random and stationary time series. Furthermore this estimator is unbiased. As estimator with an increasing scaling behaviour it overcomes the estimation problem (E1) from section \ref{sec:acfintro}. Below we will see that nonstationarity and therefore the estimation problem (E2) is still a problem.

\subsection{Nonstationary time series}
In the following two subsections we calculate $\langle
\widehat{R^2}_\text{seg}(s)\rangle$ for two different types of nonstationarity
and study the bias of the estimator. We have to analyze the
autocovariance function of the time series in the $\nu$-th segment $\langle
x(i+d_\nu)x(j+d_\nu) \rangle$ of the increment representation in
Eq.~(\ref{calku}). In the path representation of Eq.~(\ref{msdestiii}) it is
immediately clear that the estimator fails because it is averaged over
nonstationary displacements. Nevertheless we analyze
$\langle\widehat{R^2}_\text{seg}(s)\rangle$ in the increment representation
because first it shows clearly why this estimator fails and second it allows
modification to overcome the problem of nonstationarity. 

\subsubsection{Intrinsic nonstationarity}
If a time series is represented
by a nonstationary stochastic process without additional trends then this
nonstationarity is called intrinsic, see \cite{hoell3}. As example we
investigate Brownian motion and show that the estimator
$\widehat{R^2}_\text{seg}(s)$ is biased. \\ \\ 
  In order to calculate the mean of the estimator of Eq.~(\ref{calku}) we need
  first to calculate the autocovariance function of the time series in the
  $\nu$-th segment $\langle x(i+d_\nu)x(j+d_\nu) \rangle$ which is here
  identical to the autocovariance function of Brownian motion in the $\nu$-th
  segment $\langle \epsilon(i+d_\nu)\epsilon(j+d_\nu) \rangle$. The
  autocovariance function can be split into two parts 
\begin{equation}\begin{split}\label{firstsplitti}
\langle \epsilon(i+d_\nu)\epsilon(j+d_\nu) \rangle &= \text{min}(i+d_\nu,j+d_\nu)\\
&=\text{Cov}(i,j)+d_\nu
\end{split}\end{equation}
with $\text{Cov}(i,j)=\text{min}(i,j)$. Hence the autocovariance function in
$\nu$-th  segment is the autocovariance function of the first segment plus 
the shift factor. Interestingly, the dependence of the segment $\nu$ is
fully described by the second part, see figure \ref{fig:04}. This splitting of
the autocovariance function leads also to a splitting of the mean of the
estimator in Eq.~(\ref{calku}), namely 
\begin{equation}
\langle\widehat{R^2}_\text{seg}(s)\rangle  =  \sum_{i,j=1}^K \text{Cov}(i,j) + \frac{1}{K} \sum_{\nu=1}^K \sum_{i,j=1}^s d_\nu
\end{equation}
where the segment average $1/K\sum_{\nu=1}^K$ gives $1$ for the first part on
the right hand side because $\text{Cov}(i,j)$ is independent of the segment
$\nu$. This first part is exactly the path MSD $R^2(s)$. Therefore the second
part is the bias of the estimator 
\begin{equation}
B(s) = \frac{1}{K} \sum_{\nu=1}^K \sum_{i,j=1}^s d_\nu.
\end{equation}
So the estimator $\widehat{R^2}_\text{seg}(s)$ is biased. And the bias is the
reason why the estimator $\widehat{R^2}_\text{seg}(s)$ cannot detect the
scaling of the path. This can be shown by calculating $R^2(s)$ and $B^2(s)$
explicitely. The path MSD is calculated as 
\begin{equation}
R^2(s) = \frac{1}{3}s^3+ \frac{1}{2}s^2+\frac{1}{6}s
\end{equation}
which scales cubically
\begin{equation}
R^2(s) \sim \frac{1}{3} s^3
\end{equation}
for large $s$ and therefore gives a fluctuation parameter $\alpha=3/2$ as
discussed in section \ref{sec:msd}. The bias $B(s)$ can be calculated for
subseguent shifting of the segments with the shift factor $d_\nu=v-1$ and
$K=N-s+1$ segments and also for disjoint shifting with the shift factor
$d_\nu=(\nu-1)s$ and $K=N/s$ segments. Note that we use $K=N/s$ and not
$K=\lfloor N/s \rfloor$ for the sake of simplicity. In both cases we obtain 
\begin{equation}
B(s) = - \frac{1}{2}s^3+\frac{N}{2}s^2.
\end{equation}
The segment length $s$ cannot be larger than the time axis $N$. Even more, the
segmentwise averaging procedure requires enough segments $K$ in order to obtain a
reliable estimation of the autocovariance function. This requirement reduces
the largest possible value of $s$ to some value smaller than $N$. 
For those allowed values of $s$ the scaling of the bias $B(s)$ dominated
by the quadratic term 
\begin{equation}
B(s) \sim \frac{N}{2} s^2
\end{equation}
because the prefactor of the quadtratic term, $N/2$, is bigger than $s$ and
hence $\frac{N}{2} s^2 > \frac{1}{2}s^3$. And since
$\langle\widehat{R^2}_\text{seg}(s)\rangle$ is the sum of $R^2$ and $B^2$ we
obtain quadratic scaling of the mean of the estimator 
\begin{equation}
\langle\widehat{R^2}_\text{seg}(s)\rangle \sim \frac{N+1}{2} s^2
\end{equation}
with the same argument as for the scaling of $B^2$. This means that the
estimator estimator $\widehat{R^2}_\text{seg}(s)$ applied on a time series
which is a single realisation of Brownian motion estimates a fluctuation
parameter of 
\begin{equation}
\hat{\alpha}=1.
\end{equation}
Here $\hat{\alpha}$ is the estimated slope of $\widehat{R^2}_\text{seg}(s)$ in
the log-log plot which can be observed after applying the estimator
$\widehat{R^2}_\text{seg}(s)$ on a single realisation of Brownian motion. This
estimated fluctuation parameter $\hat{\alpha}=1$ is a known result for the
method of fluctuation analysis \cite{hoell3} and obviously a flaw of this
method. In fact it is a wrong estimation because for Brownian motion the
fluctuation parameter is $\alpha=3/2$. The reason for this wrong scaling
comes from the bias $B^2(s)$ which dominates the scaling of
the estimator. In addition, we also observe numerically the same scaling
$\hat{\alpha}=1$ for FBM with $1/2<H<1$ and assume similar behaviour of the
bias. In summary, in the case of a pure random and nonstationary time series
as in the case of BM or FBM the estimator of the path MSD using segmentwise
averaging procedure has a different scaling than the path MSD. 

\begin{figure}
  \includegraphics[width=1\linewidth]{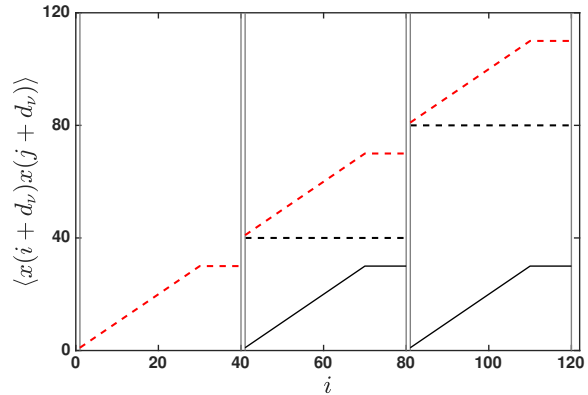}
  \caption{\small{Splitting of the autocovariance function in the $\nu$-th
      segment $\langle x(i+d_\nu)x(j+d_\nu)\rangle=\langle
      \epsilon(i+d_\nu)\epsilon(j+d_\nu)\rangle$ (dashed red line) of Brownian
      motion into the autocovariance function of the first segment
      $\text{Cov}(i,j)$ (solid black line) and the autocovariance difference
      $d_\nu$ (dashed black line), see Eq.~(\ref{firstsplitti}). Here it is
      $j=30$. And the segments have a length of $s=40$ and are disjoint with
      shift factor $d_\nu=(\nu-1)s$.}} 
  \label{fig:04}
\end{figure}

\subsubsection{External nonstationarity}
Here we calculate $\langle \widehat{R^2}_\text{seg}(s)\rangle$ for a time
series with an external nonstationarity. We model this by a stationary
stochastic process $\epsilon(t)$ with an additional deterministic trend, see
\cite{hoell3}. 
As example we investigate white noise added to a linear trend,
\begin{equation}
x(t) = \epsilon(t) + m t
\end{equation}
and show that the estimator $\widehat{R^2}_\text{seg}(s)$ is biased. The
autocovariance function of the time series in the $\nu$-th segment can again be
split into two parts 
\begin{equation}\label{splitmit}
\langle x(i+d_\nu)x(j+d_\nu)\rangle = \text{Cov}(i,j) + (i+d_\nu)(j+d_\nu)
\end{equation}
with the autocovariance function of white noise $\text{Cov}(i,j)=1$ for $i=j$
and $0$ for $i\ne j$. The second part on the right hand side comes from the
linear trend. This part completely governs the dependence of the
segments. A similar splitting of $\langle x(i+d_\nu)x(j+d_\nu)\rangle$ has
also been found for the intrinsic nonstationarity of Brownian motion where the
first part is the autocovariance function in the first segment and the second
part describes the dependence of the segments. The following discussion will
therefore be similar to the Brownian motion case.\\ \\ 
The splitting of the autocovariance function of Eq.~(\ref{splitmit}) leads
also to a splitting of the mean of estimator in Eq.~(\ref{calku}), namely 
\begin{equation}
\langle\widehat{R^2}_\text{seg}(s)\rangle  =  \sum_{i,j=1}^s \text{Cov}(i,j) + \frac{m^2}{K} \sum_{\nu=1}^K\sum_{i,j=1}^s (i+d_\nu)(j+d_\nu).
\end{equation}
Again the first part is the path MSD and therefore the second part is the bias of the estimator
\begin{equation}
B(s) = \frac{m^2}{K} \sum_{\nu=1}^K \sum_{i,j=1}^s (i+d_\nu)(j+d_\nu).
\end{equation}
So the estimator is biased. The path MSD of white noise is given by
\begin{equation}
R^2(s) = s
\end{equation}
and therefore gives a fluctuation parameter $\alpha=1/2$ as discussed in
section \ref{sec:msd}. The full solution of the bias $B(s)$ depends on the
segmentation. For subsequent segmentation the bias is 
\begin{equation}
B(s) = m^2\left(\frac{1}{12}s^4-\frac{N+1}{6} s^3 + \frac{4N^2+8N+3}{12}s^2\right)
\end{equation}
and for disjoint segmentation the bias is
\begin{equation}
B(s) = -\frac{m^2}{12}s^4 + \frac{4N^2+6N+3}{12}m^2s^2.
\end{equation}
Nevertheless for allowed values of $s$ as explained for Brownian motion both solutions are approximately similar. The bias scales quadratically
\begin{equation}
B(s) \sim \frac{N^2m^2}{3}s^2
\end{equation}
because of the largest prefactor $N^2/3$. And since
$\langle\widehat{R^2}_\text{seg}(s)\rangle$ is the sum of $R^2$ and $B$ we
obtain quadratic scaling of the mean of the estimator 
\begin{equation}
\langle\widehat{R^2}_\text{seg}(s)\rangle \sim \frac{N^2m^2}{3} s^2
\end{equation}
with the same argument as for the bias. This means that the estimator
$\widehat{R^2}_\text{seg}(s)$ applied to a time series which is a
composition of white noise and a linear trend asymtotically estimates a fluctuation
parameter of 
\begin{equation}
\hat{\alpha}=1.
\end{equation}
So the estimator is not able to detect the fluctuation parameter $\alpha=1/2$
of white noise. Numerical tests also show $\hat{\alpha}=1$ where we use FGN
with $\alpha=H$ and $1/2<H<1$ instead of white noise. And we obtain also
$\hat{\alpha}=1$ when we use trends of higher order in $t$ instead of linear. The
bias which comes here only from the additive trend dominates the scaling of
the estimator and destroys information about the stationary stochastic
process. This wrong scaling of $\hat{\alpha}=1$ is well-known for the method
of fluctuation analysis as it can be seen in \cite{kantelhardt2} but has not
yet been explained by the influence of the bias on the scaling behaviour.\\ \\

\begin{figure}
  \includegraphics[width=1\linewidth]{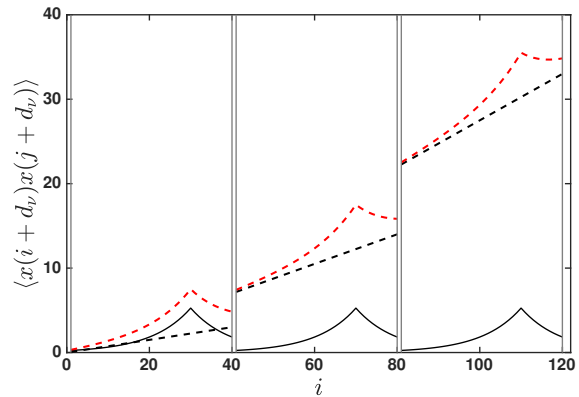}
  \caption{\small{Splitting of the autocovariance function in the $\nu$-th segment $\langle x(i+d_\nu)x(j+d_\nu)\rangle$ (dashed red line) of an AR($1$) process with additive linear trend with slope $0.05$ into the autocovariance function of the first segment of the AR($1$) process $\text{Cov}(i,j)$ (solid black line) and the autocovariance difference $0.05^2(i+d_\nu)(j+d_\nu)$ (dashed black line), see Eq.~(\ref{splitmit}). The AR($1$) process has the autocorrelation function $C(|i-j|)=0.9^{|i-j|}$.  Here it is $j=30$. And the segments have a length of $s=40$ and are disjoint with shift factor $d_\nu=(\nu-1)s$. Note that we use white noise instead of an AR($1$) process in the text of this section. But the situation of the splitting of the autocovariance function remains.}}
  \label{fig:05}
\end{figure}

\subsubsection{Summary}
In the previous subsections we have seen that the estimator
$\widehat{R^2}_\text{seg}(s)$ yields an estimation of the scaling exponent of
$\widehat{\alpha}=1$ in the case of nonstationarity no matter what stochastic
process is underlying the time series. The estimator is neither able to detect
the fluctuation parameter $\alpha$ of the path for intrinsic nor for external
nonstationary time series. Here we summarize these results, which are
the motivation for detrending methods in the next section. \\ \\ 
The mean of the estimator $\langle \widehat{R^2}_\text{seg}(s)\rangle$ depends on the autocovariance function of the time series in the $\nu$-th segment. For intrinsic and external nonstationarity this can be split into the autocovariance function of the stochastic process in the first segment and a segment-dependent part
\begin{equation}
\langle x(i+d_\nu)x(j+d_\nu) \rangle = \text{Cov}(i,j) + D_\nu(i,j).
\end{equation}
We call from now $D_\nu(i,j)$ the autocovariance difference of the time series
$x(t)$. This splitting leads to a splitting of the mean of the estimator
\begin{equation}
\langle \widehat{R^2}_\text{seg}(s)\rangle =R^2(s) + B(s),
\end{equation}
namely into the path MSD and the bias
\begin{equation}
B(s) = \frac{1}{K}\sum_{\nu=1}^K \sum_{i,j=1}^s D_\nu(i,j).
\end{equation}
The bias can be seen as the estimator applied to the autocovariance
differences. Due to the limitation $s<N$, the bias 
itself is dominated by one summand which scales like $s^2$, which then
dominates the scaling 
of the mean of the estimator
\begin{equation}
\langle \widehat{R^2}_\text{seg}(s)\rangle \sim B(s) \sim s^2\;.
\end{equation}
This leads to an estimation of the fluctuation parameter of
$\widehat{\alpha}=1$ for a given time series no matter what stochastic process
is realised in this time series. \\ \\ 
In contrast, for stationary time series the autocovariance differences are
zero and so is the bias. In that case the estimator is unbiased. 
 We will provide in the
following an estimator for the path MSD which is unbiased also in the case of
nonstationarity which is therefore a solution to the estimation problem (E2)
of section \ref{sec:acfintro}.

\section{Detrending methods}\label{sec:dobm}
\subsection{Motivation}
Given a time series we want to estimate the scaling behaviour of the path MSD
of its stochastic component. For this task the estimator
$\widehat{R^2}_\text{seg}(s)$ is unsuitable due to its failure for
nonstationary time series. This can be easily understood in its path
representation
$\widehat{R^2}_\text{seg}(s)=1/K\sum_{\nu=1}^K(y(s+d_\nu)-y(d_\nu))^2$ because
the displacements $y(s+d_\nu)-y(d_\nu)$ are nonstationary. But we used the
increment representation
$\widehat{R^2}_\text{seg}(s)=1/K\sum_{\nu=1}^K\sum_{i,j=1}^sx(i+d_\nu)x(j+d_\nu)$
of the estimator to show in detail the origin of a bias and how this bias destroys the
asymptotic scaling. To overcome this problem we introduce here a bunch of new
quantities which are all analogue to the already investigated ones. Those
new quantities are the foundation of detrending methods. 
\subsection{The fluctuation function}
For the following we want to recall that the path was defined as the
cumulative sum of the stochastic process $r(t)=\sum_{i=1}^t\epsilon(i).$
Everything what follows now is based on the replacement of the squared path displacement in the increment representation
$(r(s)-r(0))^2=\sum_{i,j=1}^s\epsilon(i)\epsilon(j)$ with the so-called
generalised squared path displacement 
\begin{equation}
f^2(s) = \sum_{i,j=1}^s \epsilon(i)\epsilon(j)L(i,j,s)
\end{equation}
which is weighted in the increment representation with the weights
$L(i,j,s)$. The purpose of the weights $L(i,j,s)$ is to suppress the effects of
 external nonstationarity and to guarantee the correct scaling with $\alpha>1$
 for intrinsic nonstationarities. The next important new quantity is a
generalistation of the path MSD. The path MSD was defined as the mean of the
squared path displacement $R^2=\langle (r(s)-r(0))^2 \rangle$. We now define
the fluctuation function as mean of the generalised squared path displacement 
\begin{equation}
F^2(s) = \langle f^2(s)\rangle.
\end{equation}
Using the definition of the generalised squared path displacement the fluctuation function reads
\begin{equation}\label{fdef}
F^2(s) = \sum_{i,j=1}^s \text{Cov}(i,j) L(i,j,s).
\end{equation}
for which $R^2(s)=\sum_{i,j=1}^s\text{Cov}(i,j)$ of Eq.~(\ref{fullmsd}) is the analogue equation of the path MSD.

\subsection{Estimator of the fluctuation function}
We introduce the estimator of the fluctuation function
 \begin{equation}\begin{split}\label{flucweight}
\widehat{F^2}(s) &=\sum_{i,j=1}^s \widehat{\text{Cov}}_\text{seg}(i,j) L(i,j,s) \\
&=\frac{1}{K} \sum_{\nu=1}^K \sum_{i,j=1}^s x(i+d_\nu)x(j+d_\nu)L(i,j,s)
\end{split}\end{equation}
which is similarly constructed as $\widehat{R}^2_\text{seg}(s)$ of Eq.~(\ref{defestr2}). This means we have replaced the autocovariance function in Eq.~(\ref{fdef}) by the segmentwise averaging estimator of the autocovariance function. Eq.~(\ref{flucweight}) implies that the generalised squared path displacement $f$ is estimated in the $\nu$-th segment by
\begin{equation}\label{detrdispl}
f_\nu^2(s) = \sum_{i,j=1}^s x(i+d_\nu)x(j+d_\nu)L(i,j,s)
\end{equation}
which is analogue to the squared profile displacement in the increment representation $(y(s+d_\nu)-y(d_\nu))^2=\sum_{i,j=1}^s x(i+d_\nu)x(j+d_\nu)$. Hence the estimator of the fluctuation function can also be understood as average of these estimators of the generalised squared path displacements
\begin{equation}\label{flucpath}
\widehat{F^2}(s) = \frac{1}{K} \sum_{\nu=1}^K f^2_\nu(s)
\end{equation}
averaged over all segments. This equation can be seen as the path representation of the estimator. We should note that in the existing literature the term "fluctuation function" is used for the estimator of the fluctuation function. But in our framework we need to be more careful with the concepts.
\subsection{Basic principles of detrending methods}
A detrending method provides a way to specify weights $L(i,j,s)$ such that the
fluctuation function respectively its estimator fulfil the following two principles:
\begin{itemize}
\item[(L1)] \underline{Scaling:} The fluctuation function should have the same asymptotic scaling behaviour as the path MSD
\begin{equation}
F^2(s) \sim s^{2\alpha}
\end{equation}
with the fluctuation parameter $\alpha$.
\item[(L2)] \underline{Unbiasedness:} The estimator of the fluctuation function should be unbiased
\begin{equation}
\langle \widehat{F^2}(s) \rangle = F^2(s).
\end{equation}
\end{itemize}
The principles (L1) and (L2) solve the estimation problems (E1) and (E2). The
first principle (L1) means that the fluctuation function scales increasingly
with two times the fluctuation parameter $\alpha$. So the weights should have no
influence on the scaling behaviour compared to the path MSD. With such an
increasing power law the estimation problem (E1) is overcome. Of course the
same scaling should be exhibited by the estimator which is governed by the second
principle. Taking the mean of the estimator $\langle \widehat{F^2}(s) \rangle$
in Eq.~(\ref{flucweight}) creates the autocovariance function $\langle
x(i+d_\nu)x(j+d_\nu) \rangle$. For the above explained types of
nonstationarity it splits into the autocovariance function $\text{Cov}(i,j)$
and the autocovariance differences $D_\nu(i,j)$. The second principle
which requires a bias of zero then leads to  
\begin{equation}\label{biasfluc}
B(s) =\frac{1}{K} \sum_{\nu=1}^K \sum_{i,j=1}^s D_\nu(i,j)L(i,j,s)=0.
\end{equation}
This equation is the mathematical formulation of what detrending should
achieve. In the literature, detrending is used to describe methods which, by 
removing non-stationarities from (parts of) the time series $x(t)$, restore
the correct scaling.
In our formalism, detrending means that the
differences of the autocovariance function between segments has no influence
on the estimation of the fluctuation function. In addition, for stationary
processes (L2) holds trivially because the autocovariance differences are zero
$D_\nu(i,j)=0$. So the second principle (L2) indeed overcomes the estimation
problem (E2).\\ \\ 
Unfortunately it is not easy to find appropriate weights $L(i,j,s)$. But
luckily there exist already methods such as detrended fluctuation analysis and
detrending moving average which serve as possible candidates for being
examples of detrending methods. But their estimators of the
fluctuation function are not in the form of Eq.~(\ref{flucweight}) and it is
some effort necessary to show the equivalence. 

\subsection{Practical implementation}\label{sec:practical}
Let us assume we know the specific form of the weights $L(i,j,s)$. Given a time series $x(t)_{t=1}^N$ the implementation of detrending methods consists of three steps:
\begin{itemize}
\item[(1)] We divide the time axis into $K$ segments of length $s$ with the $\nu$-th segment given by $[1+d_\nu,s+d_\nu]$.
\item[(2)] In every segment we calculate the estimator of the generalised squared profile displacements $f_\nu^2(s)$.
\item[(3)] We average $f^2_\nu(s)$ over all segments and obtain the estimator of the fluctuation function $\widehat{F^2}(s)$.
\end{itemize}
All three steps are repeated for different values of the segment length $s$. Because of the scaling behaviour
\begin{equation}
\widehat{F^2}(s) \sim s^{2\widehat{\alpha}}.
\end{equation}
we find an estimation of the fluctuation parameter $\widehat{\alpha}$ by linear fitting in the log-log plot.

\subsection{Detrending procedure}
The detrending procedure is the centerpiece of detrending methods. And since we provided a new and general framework we want to summarize what we now understand under detrending. The following three descriptions are used equivalently:
\begin{itemize}
\item[(A)] When a time series is successfully detrended in the framework of detrending methods then the estimator of the fluctuation function scales asymptotically as the path MSD. Hence we can observe the fluctuation parameter $\alpha$ of the stochastic process underlying the time series.
\item[(B)] The principle (L2) holds which means that the estimator of the fluctuation function is unbiased. This implies a zero bias $B(s)=0$ which is the case when the autocovariance differences between segments $D_\nu(i,j)$ have no influence on the estimation procedure.
\item[(C)] The estimators of the generalised squared path displacements $f_\nu^2(s)$ are identically distributed with respect to the segments. Then the average of $f_\nu^2(s)$ over all segments provides a unbiased estimation of the fluctuation function.
\end{itemize}
Description (A) is what is usually understood as detrending in the literature. For instance, the fluctuation parameter $\alpha$ of the DFA fluctuation function for FBM has been analytically derived in \cite{movahed}. There the original description of the fluctuation function  which is not the increment representation has been used. Hence this derivation is unaware of the unbiasedness of the estimator of the fluctuation function which is exactly stated in description (B) and which is firstly described here in this article. Description (C) has been has already been used in \cite{hoell3} for DFA applied on FBM. There it has been pointed out that the statistical equivalence of the estimator of the generalised squared path displacements
\begin{equation}\label{yu}
\langle f_\nu^2(s)\rangle = \langle f_\omega^2(s)\rangle
\end{equation}
for all segments $\nu$ and $\omega$ is necessary in order to satisfy the first principle (L1) of detrending methods. There the unbiasedness of the estimator of the fluctuation function was not mentioned explicitily. But in the here presented framework Eq.~(\ref{yu}) is exactly the second principle (L2).

\subsection{Superposition principle}
The case of external nonstationarity served as motivation to establish detrending methods. It consists of a stationary stochastic process and a deterministic trend. Actually this is a special case of two general processes. If the time series is a composition of two processes $x(t)=x_1(t) + x_2(t)$ and both are independent of each other then the fluctuation function is the sum of two single ones
\begin{equation}
F^2(s)  =  F_1^2(s)  + F_2^2(s)
\end{equation}
where $F^2_n(s)$ is the fluctuation function of the process $\{x_n(t)\}_{t=1}^N$ with $n=1$ and $2$, see \cite{hu}. This is the superposition principle. Independence between both single processes implies zero cross-correlation $\langle x_1(i)x_2(j) \rangle=0$ for all $i$ and $j$. Using this in Eq.~(\ref{fdef}) gives the superposition principle of the fluctuation function. When the second principle (L2) holds then the superposition principle also holds for the estimators of the fluctuation functions. \\ \\
As special case of external nonstationarity, when the first process is a stationary stochastic process and the second process is a deterministic trend then the superposition principle yields that $F_1^2(s)$ is the full fluctuation function and $F_2^2(s)$ is the bias which should be zero under successfull detrending.

\subsection{Factorisable weights and wavelets}
If the weights are factorisable
\begin{equation}\label{factorise}
L(i,j,s) = l(i,s) l(j,s)
\end{equation}
then the estimator of the generalised path displacement can be written as weighted sum of the time series
\begin{equation}\label{also-wavelet}
f_\nu(s) =  \sum_{i=1}^s x(i+d_\nu) l(i,s)
\end{equation}
with weights $l(i,s)$. This equation is a generalisation of the relationship
between the profile displacement and the time series as $y(s+d_\nu)-y(d_\nu) =
\sum_{i=1}^s x(i+d_\nu)$. As we will see later the method of detrending moving
average DMA has factorisable weights, but detrended fluctuation analysis DFA
does not. \\ \\
Another popular estimation method for the scaling exponent with build-in
detrending relies on a wavelet transform \cite{abry,veitch}.
The quantity which corresponds to the fluctuation function Eq.~(\ref{fdef}) is
there the time average of the squared scale-$s$ wavelet coefficient, averaged
over all disjoint windows of length $s$ contained in the time series. Its
scaling exponent $\beta$ is related to the exponent $\alpha$ used here  by $\beta=2\alpha-1$. The
detrending is here achieved by choosing wavelets whose first $n$ moments
vanish, so that power law trends up to order $n-1$ are projected out from the
wavelet transform. The method becomes particularly simple if wavelets are
taken from the famility of Haar wavelets \cite{abry}. The wavelet method therefore
is a variant of Eq.~(\ref{also-wavelet}), where the wavelet is a special choice of 
the kernel $l(i,s)$, see also \cite{kiyono5}.

\section{DFA and DMA}
In previous section, we introduced basic principles of detrending methods. Now
we will show that detrended fluctuation analysis (DFA) and detrending moving
average (DMA) are examples of such methods. In order to do so we need to
answer two questions: How can we come from the original form of the estimator
of the fluctuation function of DFA and DMA to the here introduced one in the
increment representation? And what are the specific expressions of the weights
$L(i,j,s)$ for DFA and DMA? Both methods are constructed differently
nevertheless we present them simultaneously in the following to emphasize
their similarities and differences. We then show the ability of DFA and DMA to
fulfill basic principles (L1) and (L2). 
\subsection{Original description}
We present here the original description of DFA \cite{kantelhardt2} and DMA
\cite{carbone4} as it is applied as a tool on real time series. Here we only
investigate centered DMA and not backward and forward DMA. In their original 
definitions, fluctuation functions are not described in the
increment representation as in Eq.~(\ref{flucweight}). Actually it is not
evident if and how these original forms can be related to our
description. \\ \\ 
Given is a time series $\{x[t]\}_{t=1}^N$. For both methods DMA and DFA the 3
steps of the practical implementation of section \ref{sec:practical} are
detailed as follows: 
\begin{itemize}
\item[(1)] We divide the time axis into $K$ segments of length $s$ with the
  $\nu$-th segment given by $[1+d_\nu,s+d_\nu]$. For DMA the segments are
  shifted by one time point so that $d_{\nu,\text{DMA}}=\nu-1$ and
  $K_\text{DMA}=N-s+1$. For DFA the segments are disjoint so that
  $d_{\nu,\text{DFA}}=(\nu-1)s$ and $K_\text{DFA}=\lfloor N/s \rfloor$. 
\item[(2)] First we calculate in every segment $\nu$ the fitting polynomial
  $\{p_\nu^{(q)}(t)\}_{1+d_\nu}^{s+d_\nu}$ of the profile
  $\{y(t)\}_{1+d_\nu}^{s+d_\nu}$ using method of least squares. The fit can
  have any order $q \in \mathbb{N}$. The estimator of the generalised squared
  path displacement for DMA is given by the squared distance between the
  profile and the fit at the middle point $\nu+(s-1)/2$ of the segment 
\begin{equation}\begin{split}\label{vardma}
f^2_{\nu,\text{DMA$q$}}(s)=\Bigg(&y\left(\frac{s+1}{2}+d_{\nu,\text{DMA}}\right)\\
-&p_\nu^{(q)}\left(\frac{s+1}{2}+d_{\nu,\text{DMA}}\right)\Bigg)^2.
\end{split}\end{equation}
We only allow odd $s$ for DMA. For DFA it is the averaged squared distance between profile and fit
\begin{equation}\begin{split}\label{vardfa}
f^2_{\nu,\text{DFA$q$}}(s)=\frac{1}{s} \sum_{t=1}^s\Big(&y(t+d_{\nu,\text{DFA}})\\
-&p_\nu^{(q)}(t+d_{\nu,\text{DFA}})\Big)^2
\end{split}\end{equation}
averaged over all points in the segment. The order of the detrending
polynomial $q$ is usually indicated by calling the methods DMA$q$ and DFA$q$.
\item[(3)] We average $f^2_\nu(s)$ over all segments and obtain the estimator of the fluctuation function $\widehat{F^2}(s)$.
\end{itemize}
All three steps are repeated for different values of the segment length $s$
and then we ideally observe the scaling behaviour $\widehat{F^2}(s) \sim
s^{2\widehat{\alpha}}$ where $\widehat{\alpha}$ is an estimation of the
fluctuation parameter $\alpha$. We will now work out how this
 original definition of the estimator of the fluctuation function
for DFA and DMA is related to properties of the stochastic process
$\epsilon(t)$, especially to its correlation structure and (non-)stationarity. 

\subsection{Increment representation}\label{sec:acovreprdfa}
The estimators of the generalised squared path displacements $f_\nu^2$ of DMA
in Eq.~(\ref{vardma}) and DFA in Eq.~(\ref{vardfa}) are not written in the
increment representation $f_\nu^2=\sum_{i,j=1}^s
x(i+d_\nu)x(j+d_\nu)L(i,j,s)$. Therefore the weights $L(i,j,s)$ are
unknown. The starting point of transforming the original definition into the
increment representation is to write the residual
$y(t+d_\nu)-p_\nu^{(q)}(t+d_\nu)$ as a function of the time series elements $x(t)$. We show in
appendix \ref{sec:residual} that the residual in the $\nu$-th segment can be
expressed as weighted sum of the time series 
\begin{equation}\label{weightedsum}
y(t+d_\nu)-p_v^{(q)}(t+d_\nu)=\sum_{i=1}^s x(i+d_\nu) \omega^{(q)}(i,t)
\end{equation}
with $t \in [1,s]$ for DFA and $t=(s+1)/2$ for DMA. An important observation
is that the dependence on the segment $\nu$ only occurs as argument in the
time series but not the weights. The weights are explicitly given as 
\begin{equation}\label{jkl}
\omega^{(q)}(i,t)=\Theta(t-i)-\sum_{m=0}^q t^m \sum_{n=0}^q  \left(\mathbb{S}^{-1}\right)_{m+1,n+1} \sum_{k=i}^s k^n,
\end{equation}
see figure \ref{fig:06} and \ref{fig:07}. The Heaviside function $\Theta$
comes from the profile and the second part from the polynomial fit. The
$(q+1)\times(q+1)$ matrix $\mathbb{S}$ has the matrix elements $S_{mn} =
\sum_{k=1}^s k^{m+n-2}$. In principal the second part can be explicitly
calculated for any order of detrending $q$. For the lowest order of detrending
$q=0$ the weight is 
\begin{equation}
\omega^{(0)}(i,t) = \Theta(t-i) - \frac{s-i+1}{s}.
\end{equation}
We provide a Mathematica code in appendix
\ref{sec:mathematica}.\ref{sec:appwei} which calculates the weights $\omega^{(q)}(i,t)$
for specific but arbitrary order of detrending $q$. With
Eq. (\ref{weightedsum}) we can write $f_\nu^2(s)$ of Eq. (\ref{vardma}) and
(\ref{vardfa}) in the increment representation. This is obtained by first
inserting the weighted sum of Eq. (\ref{weightedsum}) in Eq. (\ref{vardma})
and (\ref{vardfa}) and then expanding the square. Then we simply can read
the weights. For DMA it is the product 
\begin{equation}\label{dmakern}
L_\text{DMA$q$}(i,j,s)=\omega^{(q)}\left(i,\frac{s+1}{2}\right)\omega^{(q)}\left(j,\frac{s+1}{2} \right),
\end{equation}
see appendix \ref{sec:flucfunwe}.\ref{sec:flucfunwedma} and for DFA it is the average of the products
\begin{equation}\label{dfakern}
L_\text{DFA$q$}(i,j,s)=\frac{1}{s} \sum_{t=1}^s \omega^{(q)}(i,t)   \omega^{(q)}(j,t),
\end{equation}
see appendix \ref{sec:flucfunwe}.\ref{sec:flucfunwedfa}. We provide a
Mathematica code in appendix \ref{sec:mathematica}.\ref{weightsflucmat} for
the weights of DFA $L_\text{DFA$q$}(i,j,s)$ for any order of detrending
$q$. For lowest order of detrending $q=0$ it is 
\begin{equation}
L_\text{DFA$0$}(i,j,s)= \frac{(i-1)(s-j+1)}{s^2}.
\end{equation}
The weights of DMA are factorisable but not the weights of DFA. Nevertheless
with both weights $L(i,j,s)$ we can write the estimator of the fluctuation
functions $\widehat{F^2}(s)$ for DMA and DFA in the increment
representation. We will show in the following the ability of the weights
$L(i,j,s)$ of both methods to fulfill the basic principles (L1) and (L2) of
section \ref{sec:dobm}.

\begin{figure}[H]
  \includegraphics[width=1\linewidth]{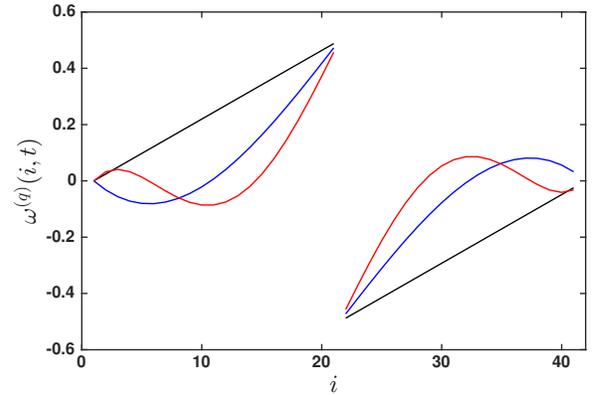}
  \caption{\small{The weights of the residual $\omega^{(q)}(i,t)$ for Eq.~(\ref{jkl}) for detrending order $q=0$ and $1$ (black line), $q=2$ and $3$ (blue line) and $q=4$ and $5$ (red line). The segment length is $s=41$ and the time point is $t=(s+1)/2=21$. Note that for $t=(s+1)/2$ the weights are identical for $q=0$ and $1$ , $2$ and $3$ as well as $4$ and $5$ but not for other $t$, see figure \ref{fig:07}.}}
  \label{fig:06}
\end{figure}

\begin{figure}[H]
  \includegraphics[width=1\linewidth]{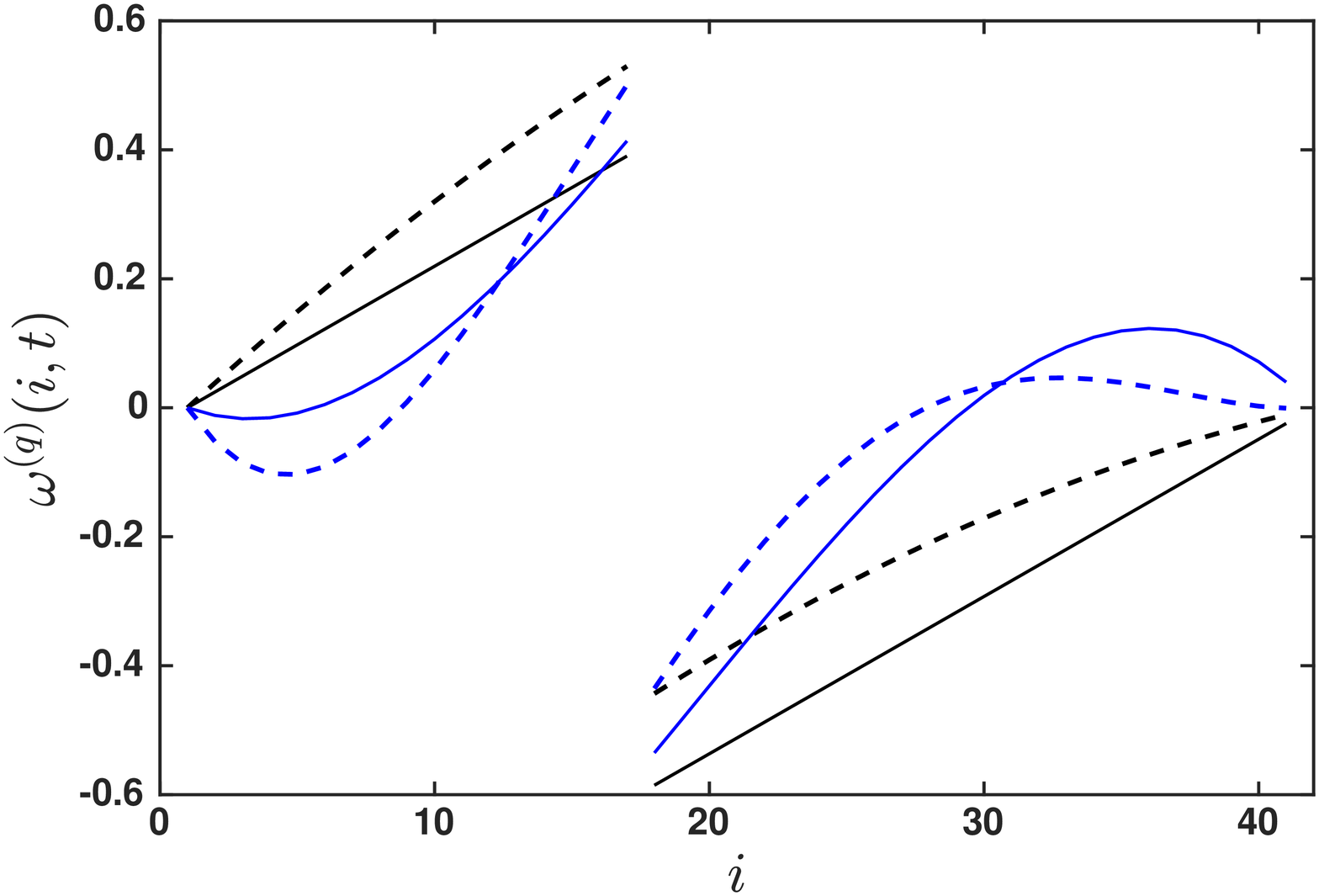}
  \caption{\small{The weights of the residual $\omega^{(q)}(i,t)$ for Eq.~(\ref{jkl}) for detrending order $q=0$ (solid black line), $1$ (dashed black line), $2$ (solid blue line) and $3$ (dashed blue line). The segment length is $s=41$ and the time point is $t=17$.}}
  \label{fig:07}
\end{figure}

\subsection{Scaling of the fluctuation function}\label{sec:scalingof}
Before we continue with our investigation we summarize 
published results about analytical derivations of the scaling behaviour of the
fluctuation function $F^2(s)\sim s^{2\alpha}$, i.e. the derivation of $\alpha$
for specific processes. For DFA using the original definition of the
fluctuation function: WN \cite{hoell} AR($1$) \cite{hoell}, FGN
\cite{taqqu2,bardet,movahed2,crato}, BM \cite{hoell} and FBM
\cite{movahed,heneghan,kiyono}. For DFA using the increment representation:
WN, AR($1$) and ARFIMA($0$,$d$,$0$) \cite{hoell2}. For DMA using the original
definition of the fluctuation function: FGN
\cite{arianos,arianos2,carbone}. Although we are not aware of any analytical
investigation of additive trends and FBM using the increment
representation, we here restrict ourselves to stationary processes.
 We present simultaneously the investigation for DFA and DMA.\\ \\ 
First we order the fluctuation function in the increment representation
$F^2(s)=\sum_{i,j=1}^s\text{Cov}(i,j)L(i,j,s)$ according to the time lag
$\tau$, namely 
\begin{equation}\begin{split}
F^2(s) &= \sum_{i=1}^s \text{Cov}(i,i)L(i,i,s) \\
&+ 2 \sum_{\tau=1}^{s-1} \sum_{i=1}^{s-\tau}\text{Cov}(i,i+\tau)L(i,i+\tau,s)
\end{split}\end{equation}
For stationary processes the autocovariance function $\text{Cov}(i,j)$ only depends on the time lag $\tau$ and not the time point $i$. Therefore we can write this equation as
 \begin{equation}\label{flucstat222}
F^2(s)=\langle x^2(i)\rangle \left( \mathcal{L}(0,s) + 2 \sum_{\tau=1}^{s-1} C(\tau)  \mathcal{L}(\tau,s)\right)
 \end{equation}
with the weights for the stationary fluctuation function
 \begin{equation}\label{statweight1}
 \mathcal{L}(\tau,s) =  \sum_{i=1}^{s-\tau} L(i,i+\tau,s),
  \end{equation}
see figure \ref{fig:08} and \ref{fig:09}. In appendix \ref{sec:statdmafluc} we
provide a detailed investigation of the weights of DMA
$\mathcal{L}_{\text{DMA}q}(\tau,s)$. We also provide a Mathematica code for
the calculation of $\mathcal{L}(\tau,s)$ for DMA in appendix
\ref{sec:mathematica}.\ref{sec:dmastatweight} and DFA in appendix
\ref{sec:mathematica}.\ref{sec:dfastatweight} for any order of detrending.  As
example we present the weight for DFA and zero order of detrending 
\begin{equation}
 \mathcal{L}_{\text{DFA}0}(\tau,s) = \frac{-\tau^3+3s\tau^2+(1-3s^2)\tau+s^3-s}{6s^2}.
\end{equation}
We also provide a Mathematica code for the calculation of $F^2(s)$ of
Eq.~(\ref{flucstat222}) for DMA in appendix
\ref{sec:mathematica}.\ref{sec:dmaflucmath} and DFA in appendix
\ref{sec:mathematica}.\ref{sec:dfaflucmath} for any order of detrending and
adjustable autocorrelation function. We tested the code for white noise,
AR($1$) and ARFIMA($0$,$d$,$0$) processes. The asymptotic scaling behaviour of
these explicit solutions is $2\alpha$ and we can therefore identify the
fluctuation parameter for large $s$.\\ \\ 
We also can understand the asymptotic behaviour of the fluctuation function
directly from Eq.~(\ref{flucstat222}), see \cite{hoell2} for the DFA
fluctuation function. The same argumentation as in \cite{hoell2} holds also
for the DMA fluctuation function as it is presented for both methods in the
following for uncorrelated and long-range correlated processes. For
uncorrelated processes as white noise the autocorrelation function is nonzero
only for $\tau=0$ and therefore the scaling is determined by the first part
$\mathcal{L}(0,s)$ of Eq.~(\ref{flucstat222}) which scales linearly 
 \begin{equation}
F^2(s)\sim  \mathcal{L}(0,s) \sim s.
 \end{equation}
 Hence the fluctuation function scales with the fluctuation parameter $\alpha=1/2$ as it is claimed by the first principle (L1). In contrast for long-range correlated processes with autocorrelation function $C(\tau) \sim \tau^{-\gamma}$ with correlation parameter $0<\gamma<1$ the sum over $\tau$ in Eq.~(\ref{flucstat222}) dominates the asymptotic scaling behaviour
 \begin{equation}
 F^2(s)\sim  \sum_{\tau=1}^{s-1} C(\tau)  \mathcal{L}(\tau,s) =  \sum_{\tau=1}^{s-1} s^{-\gamma} \mathcal{L}(\tau,s)\sim s^{2-\gamma}.
 \end{equation}
Hence the fluctuation function scales with fluctuation parameter $\alpha=1-\gamma/2$ which is also in accordance with the first principle (L1).

\begin{figure}
  \includegraphics[width=1\linewidth]{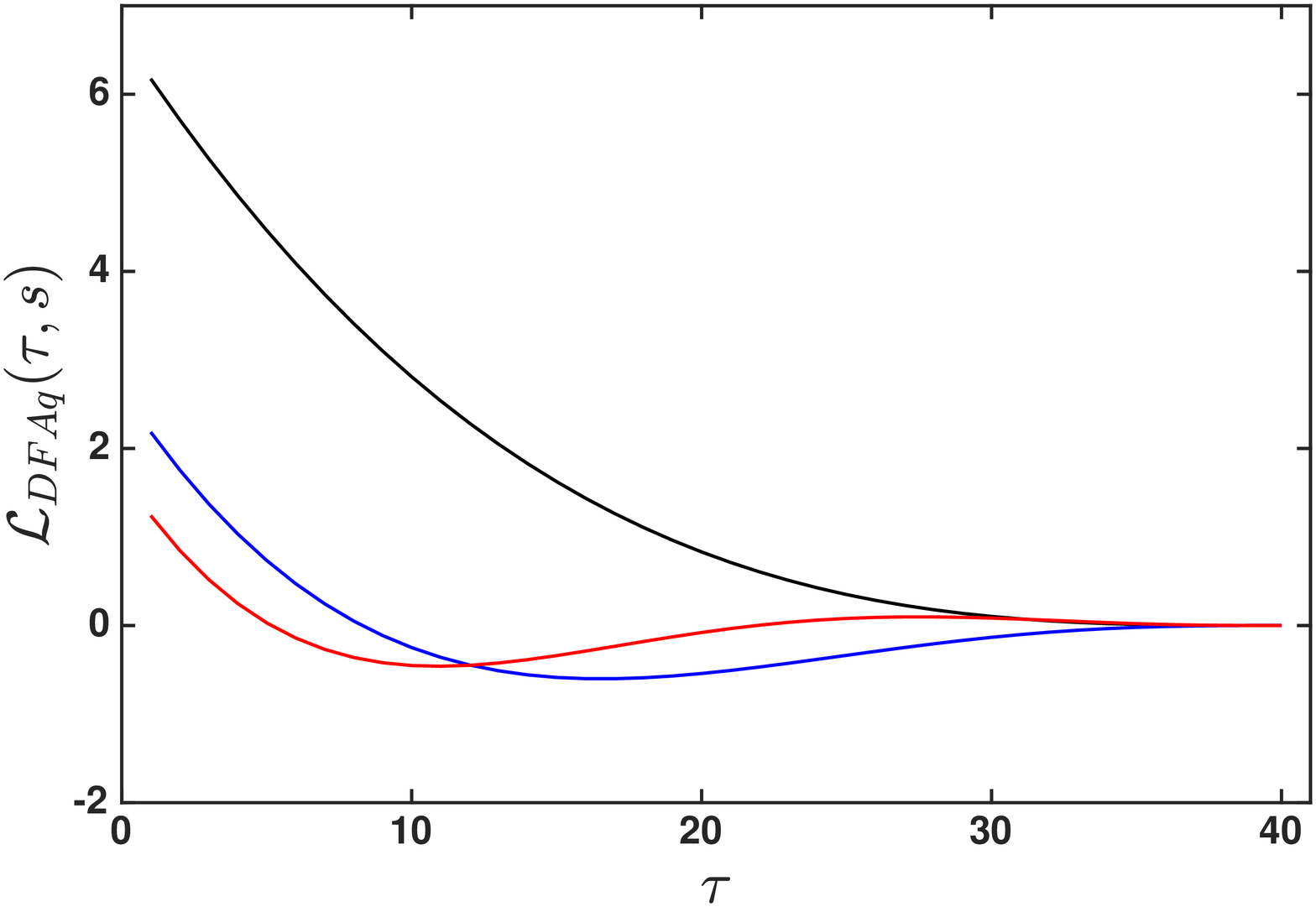}
  \caption{\small{The weights of the stationary DFA fluctuation function $\mathcal{L}_{\text{DFA}q}$ of Eq.~(\ref{statweight1}) for detrending order $q=0$ (black line), $1$ (blue line) and $2$ (red line). Here it is $s=41$.}}
  \label{fig:08}
\end{figure}

\begin{figure}
  \includegraphics[width=1\linewidth]{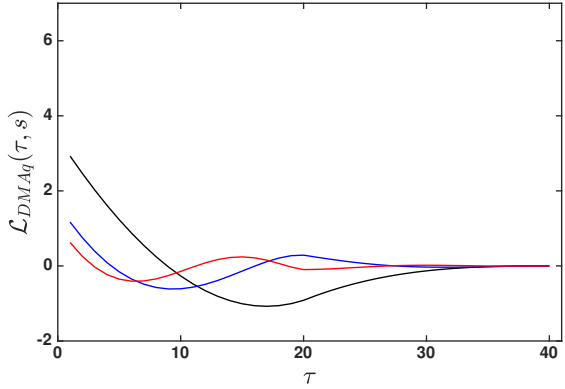}
  \caption{\small{The weights of the stationary DFA fluctuation function $\mathcal{L}_{\text{DFA}q}$ of Eq.~(\ref{statweight1}) for detrending order $q=0$ and $1$ (black line), $2$ and $3$ (blue line) as well as $4$ and $5$ (red line). Here it is $s=41$.}}
  \label{fig:09}
\end{figure}

\subsubsection{Crossover behaviour of autoregressive processes}
Stationary processes with an autoregressive part like ARMA and ARFIMA
processes show a crossover behaviour of the fluctuation function where only
asymptotically the scaling exponent of $2\alpha$ is reached. For DFA applied
on an AR($1$) process this has been investigated in \cite{hoell,maraun}. In
\cite{hoell} it has been shown that the second part $\sum_{\tau=1}^{s-1}
C(\tau)  \mathcal{L}(\tau,s)$ dominates the scaling of $F^2(s)$ for small
enough $s$ due to the nonzero and exponential decaying autocorrelation
function. In this small $s$ regime the scaling exponent is larger than $2\alpha$. Only
for large enough $s$ the scaling of the AR($1$) fluctuation function goes over
to that of the white noise case because then $\sum_{\tau=1}^{s-1}
C(\tau)  \mathcal{L}(\tau,s)$ becomes sufficiently small in comparison to
$\mathcal{L}(0,s)$. We also observe numerically a crossover behaviour for an
ARFIMA($1$,$d$,$0$) process which serves as example of a stationary long-range
correlated process with an autoregressive part. But here we cannot explain in
detail the crossover behaviour since we are not aware of a full analytical
solution of the fluctuation function because of the quite complicated form of
the autocorrelation function. \\ \\ 
Such crossover behaviour requires a relatively large amount of data in order
to observe the correct scaling. If there is not enough data then it is
difficult to distinguish between short-range and long-range correlations using
detrending methods for fluctuations \cite{meyer,hoell}.

\subsection{Unbiasedness of the estimator }
In order to check the biasedness of the estimator of the fluctuation function
we take the mean of the estimator $\langle \widehat{F^2}(s)\rangle$ and
compare the difference to the fluctuation function $F^2(s)$. $\langle
\widehat{F^2}(s)\rangle$ depends on the autocovariance function of the time
series $\langle x(i+d_\nu) x(j+d_\nu) \rangle$. If there is no difference then
the estimator is unbiased and can detect the scaling of the fluctuation
function. 
\subsubsection{Stationarity}
For stationary processes the autocovariance function of the time series
$\langle x(i+d_\nu) x(j+d_\nu) \rangle$ is independent of the shift factor
$d_v$. This means it is the autocovariance function of the stochastic process
$\langle x(i+d_\nu) x(j+d_\nu) \rangle=\text{Cov}(i,j)$ and therefore the
estimator is unbiased $\langle \widehat{F^2}(s)\rangle=F^2(s)$ and the
estimator obtains the scaling as explained in section \ref{sec:scalingof}. 
\subsubsection{Intrinsic nonstationarity}\label{sec:biasint}
As example of an intrinsic nonstationary process we investigate here FBM with the autocovariance function
\begin{equation}
\text{Cov}(i,j) = \frac{1}{2}\left(i^{2H}+j^{2H}-|i-j|^{2H} \right).
\end{equation}
This problem has already been investigated for DFA in \cite{hoell3}. But there
only the first and second segment has been compared and not all. Here we treat
all segments and also the method of DMA. The autocovariance difference between
segments $D_\nu(i,j)=\langle x(i+d_\nu) x(j+d_\nu) \rangle-\text{Cov}(i,j)$
with $\langle x(i+d_\nu) x(j+d_\nu) \rangle=\text{Cov}(i+d_\nu,j+d_\nu)$ is 
\begin{equation}\label{acovdifffbm}
D_\nu(i,j) = \frac{1}{2} \left((i+d_\nu)^{2H}+(j+d_\nu)^{2H} -i^{2H}-j^{2H} \right).
\end{equation}
We show in appendix \ref{sec:fbmdfiff} that this can be written as infinite series
\begin{equation}\label{expans}
D_\nu(i,j) = \sum_{p=0}^\infty \lambda_p (i^p+j^p)
\end{equation}
using a Taylor expansion. The prefactors $\lambda_p$ depend on $H$, $d_\nu$
and an arbitrary point where the Taylor series is evaluated at, see appendix
\ref{sec:fbmdfiff}. In \cite{hoell3} another series expansion using binomial
series for $D_2(i,j)$ with DFA shift factor $d_2=s$ was proposed. But this
series expansion does not converge always with the DMA shift factor whereas
Eq.~(\ref{expans}) does. \\ \\ 
With Eq.~(\ref{expans}) we can now check the biasedness of the estimator,
namely by taking the mean of Eq.~(\ref{flucpath}) which gives $\langle
\widehat{F^2}(s)\rangle=1/K \sum_{\nu=1}^K\langle f_\nu^2(s)\rangle$. The mean of the
estimator of the generalised squared path displacement is 
\begin{equation}\label{fbmtest3}
\langle f_\nu^2(s)\rangle = \sum_{i,j=1}^s \text{Cov}(i,j)L(i,j,s) + \sum_{i,j=1}^s D_\nu(i,j)L(i,j,s)
\end{equation}
where we used the splitting of $\langle x(i+d_\nu) x(j+d_\nu) \rangle$. With the Mathematica code for the calculation of $\langle f_\nu^2(s)\rangle$ for DMA in appendix \ref{sec:mathematica}.\ref{fv2dma} and for DFA in appendix \ref{sec:mathematica}.\ref{fv2dfa} we verified that
\begin{equation}\label{fbmtest2}
\sum_{i,j=1}^s (i^p+j^p)L(i,j,s) =0
\end{equation}
for $p\in[1,50]$ and also particular higher values of $p$. Hence it is
reasonable to assume that this equation holds for all $p$. We found that
Eq.~(\ref{fbmtest2}) is only zero for the detrending order $q\ge 1$ for DFA and
$q \ge 0$ for DMA. In those cases the estimator is unbiased $\langle
\widehat{F^2}(s)\rangle=F^2(s)$ because the bias is zero $B(s)=0$ due to
Eq.~(\ref{fbmtest2}). Finally, no matter if $\langle f_\nu^2(s)\rangle$ is
zero or not, we also provide a Mathematica code for the calcuation of $\langle
\widehat{F^2}(s)\rangle=1/K \sum_{\nu=1}^K\langle f_\nu^2(s)\rangle$ in appendix
\ref{sec:mathematica}.\ref{meanestfluc}. But due to the complicated form of
the right hand side of Eq.~(\ref{fbmtest3}) the code will not always be
successfully yielding a result.

\subsubsection{External nonstationarity}
As an example of an external nonstationarity we investigate additive trends
$\{m(t)\}_{t=1}^N$ of order $p$, i.e. $m(t)\sim t^p$. The following is similar
to the investigation of FBM from the previous section. The autocovariance
difference is 
\begin{equation}
D_\nu(i,j) = m(i+d_\nu)m(j+d_\nu).
\end{equation}
With the Mathematica code for the calculation of $\langle f_\nu^2(s)\rangle$
for DMA in appendix \ref{sec:mathematica}.\ref{fv2dma} and for DFA in appendix
\ref{sec:mathematica}.\ref{fv2dfa} we checked if the generalised squared path
displacement applied on the autocovariance differences is zero, 
\begin{equation}\label{fbmtest23}
\sum_{i,j=1}^s m(i+d_\nu)m(j+d_\nu)L(i,j,s) =0,
\end{equation}
or not. We checked orders of the trend $p\in[1,6]$. For DFA this equation is
zero if the order of detrending is $q\ge p+1$. For DMA it is $q\ge p+1$ for
odd $p$ and $q\ge p$ for even $p$, this has also been found in
\cite{carbone}. For those cases the estimator of the fluctuation function is
unbiased $\langle \widehat{F^2}(s)\rangle=F^2(s)$ because the bias is zero
$B(s)=0$ due to Eq.~(\ref{fbmtest23}).

\subsubsection{Unified picture of detrending}
For both methods DFA and DMA the detrending procedure works similarly for two
different types of nonstationarity, namely FBM and additive trends. After
segmentation of the time axis the autocovariance function of the time series
$\langle x(i+d_\nu) x(j+d_\nu) \rangle$ is different in every segment due to
the shift factor $d_\nu$. This difference is described by the autocovariance
differences $D_\nu(i,j)$. And since the estimation of the fluctuation function
depends on the product $x(i+d_\nu) x(j+d_\nu)$ it is necessary that a
successful detrending procedure gets rid of the influence of $D_\nu(i,j)$. If
that is the case then the estimator of the fluctuation function is unbiased
and therefore fulfills the second principle (L2). And exactly then the
estimator of the fluctuation function can detect the fluctuation parameter
$\alpha$ as described in the first principle (L1) even for those two different
types of nonstationarity.

\section{Summary}
We provided three main points in this article of which all of them are
similarly relevant and also linked with each other. First we motivated the
introduction of detrending methods started from basic statistical methods. We
introduced the fluctuation function as modified path MSD written in the
increment representation, i.e. the connection between the fluctuation function
and autocovariance function. Originally DFA and DMA provide an estimator of
the fluctuation function in a form for which several points are unclear: the
connection to the autocovariance function, the scaling behaviour and the
ability of treating FBM. Especially the last two points are
disadvantageous. The reason for this original description of the estimator of
the fluctuation function is that these methods evolved from prior methods. But
in the increment representation those two points can be handled which led us
to the second main point: the description of detrending methods via two basic
principles. The first principle ensures that the fluctuation function scales
asymptotically similar to the path MSD. The second principles ensures that the
estimator of the fluctuation function is unbiased. This is the centerpiece of
the detrending procedure and is carried by our construction of the fluctuation
function using a modificiation of the path MSD. Without this modification,
namely the weighting of the path MSD in the increment representation, the
estimator of the path MSD fails for nonstationary time series. And third we
showed in detail that DFA and DMA are indeed examples of detrending methods. We
explicitly verified the fulfillment of the two basic principles for both
methods. in summary, this article provides a basic overview of detrending
methods which answered fundamental questions. Furthermore we believe that this
work can serve as basis for finding advanced detrending methods depending on
specific problems in time series analysis, such as dealing with periodic
deterministic components in $x(t)$.

\begin{appendix}

\section{Residual as weighted sum}\label{sec:residual}
Here we express the residual $y(t+d_\nu)-p_\nu^{(q)}(t+d_\nu)$ of Eq.~(\ref{weightedsum}) in dependence of the time series. First we rewrite the profile and then the fit. The profile is per definition
\begin{equation}\begin{split}
y(t+d_\nu) &= \sum_{i=1}^{t+d_\nu} x(i) = \sum_{i=1}^{s+d_\nu} x(i) \Theta(t+d_\nu-i)\\
&= y(d_\nu) + \sum_{i=1+d_\nu}^{s+d_\nu} x(i) \Theta(t+d_\nu-i) \\
&=y(d_\nu) + \sum_{i=1}^s x(i+d_\nu) \Theta(t-i).
\end{split}\end{equation}
The polynomial fit in segment $\nu$ of order $q$ can be written as
\begin{equation}
p_\nu^{(q)}(t+d_\nu) = y(d_\nu) + \sum_{i=1}^s x(i+d_\nu) P_\nu^{(q)}(i,t)
\end{equation}
with the weights of the fit
\begin{equation}\label{fitweight}
P_\nu^{(q)}(i,t)= \sum_{m=0}^q (t+d_\nu)^m \sum_{n=0}^q (\mathbb{S}_\nu^{-1})_{m+1,n+1} \sum_{k=i+d_\nu}^{s+d_\nu} k^n,
\end{equation}
see \cite{hoell3}. The inverse matrix elements can be calculated as
$(\mathbb{S}_\nu^{-1})_{m+1,n+1}=(\text{adj}\mathbb{S}_\nu)_{m+1,n+1}/\text{det}\mathbb{S_\nu}$ where $\text{adj}\mathbb{S}_\nu$ is the adjugate matrix of the $(q+1)\times(q+1)$ matrix $\mathbb{S}_\nu$. The matrix $\mathbb{S}_\nu$ has the elements $S_{mn,\nu}=\sum_{i=1+d_\nu}^{s+d_\nu} i^{m+n-2}.$ Note that we wrote $\mathbb{S}=\mathbb{S}_1$ in section \ref{sec:acovreprdfa}. With the Mathematica code in appendix \ref{sec:mathematica}.\ref{sec:appwei2} we tested several orders of detrending $q$  and always found that $P_\nu^{(q)}(i,t)$ is independent of the segment $\nu$. This was tested for $d_\nu=(\nu-1)s$ and $\nu-1$. Hence we assume this independence of the segment is true for all $q$. Therefore the residual is
\begin{equation}\label{resweight}
y(t+d_\nu)-p_\nu^{(q)}(t+d_\nu) = \sum_{i=1}^s x(i+d_\nu) \omega^{(q)}(i,t)
\end{equation}
with the weights of the residual
\begin{equation}
\omega^{(q)}(i,t) = \Theta(t-i) - P_1^{(q)}(i,t).
\end{equation}

\section{$\boldsymbol{f_\nu^2(s)}$ in the increment representation}\label{sec:flucfunwe}
The estimators of the generalised squared path displacement $f_\nu^2(s)$ of DMA of  Eq.~(\ref{vardma}) and DFA of Eq.~(\ref{vardfa}) are written in their original description. Here we write them in the increment representation $f_\nu^2(s)=\sum_{i,j=1}^sx(i+d_\nu)x(j+d_\nu)L(i,j,s)$, see Eq.~(\ref{detrdispl}). We find the specific forms of the weights $L(i,j,s)$ for DMA and DFA.

\subsection{$\boldsymbol{f_\nu^2(s)}$ of DMA in the increment representation}\label{sec:flucfunwedma}
In the following we use for the middle point of the first segment the notation $\sigma=(s+1)/2$. For DMA the estimator of the generalised squared path displacement is
\begin{equation}\begin{split}
f^2_{\nu,\text{DMA$q$}}(s)=\left(y(\sigma+d_{\nu,\text{DMA}})-p_\nu^{(q)}(\sigma+d_{\nu,\text{DMA}})\right)^2,
\end{split}\end{equation}
see Eq.~(\ref{vardma}). Using the residual as weighted sum given in Eq.~(\ref{resweight}) we can write this as
\begin{equation}\begin{split}\label{ret3}
&f_{\nu,\text{DMA$q$}}^2(s)\\
 &=  \left( \sum_{i=1}^s x(i+d_{\nu,\text{DMA}}) \omega^{(q)}(i,\sigma) \right)^2, \\
&= \sum_{i,j=1}^s x(i+d_{\nu,\text{DMA}})x(j+d_{\nu,\text{DMA}})  \omega^{(q)}(i,\sigma) \omega^{(q)}(j,\sigma) .
\end{split}\end{equation}
Hence the weights of DMA are
\begin{equation}\label{gdfz}
L_{\text{DMA}q}(i,j,s) =  \omega^{(q)}(i,\sigma) \omega^{(q)}(j,\sigma)
\end{equation}
which has breaks due to the Heaviside function. In detail $f_{\nu,\text{DMA$q$}}^2(s)$ is given by
\begin{equation}\begin{split}\label{fdmadetail}
&f_{\nu,\text{DMA}q}^2(s)  \\
&= \sum_{i=1}^\sigma  \Big[ x^2(i+d_{\nu,\text{DMA}})\left(1-P_1^{(q)}(i,\sigma)\right)^2 \Big] \\
&+ \sum_{i=\sigma+1}^s \Big[ x^2(i+d_{\nu,\text{DMA}})\left(P_1^{(q)}(i,\sigma)\right)^2 \Big]\\
&+2\sum_{i=1}^\sigma\sum_{j=i+1}^\sigma \Big[ x(i+d_{\nu,\text{DMA}})x(j+d_{\nu,\text{DMA}})\\
&\hspace{1.5cm}\times\left(1-P_1^{(q)}(i,\sigma)\right)\left(1-P_1^{(q)}(j,\sigma)\right)\Big]\\
&+2\sum_{i=1}^\sigma\sum_{j=\sigma+1}^s \Big[x(i+d_{\nu,\text{DMA}})x(j+d_{\nu,\text{DMA}})\\
&\hspace{1.5cm}\times\left(1-P_1^{(q)}(i,\sigma)\right)\left(-P_1^{(q)}(j,\sigma)\right) \Big]\\
&+2\sum_{i=\sigma+1}^{s-1} \sum_{j=i+1}^s \Big[ x(i+d_{\nu,\text{DMA}})x(j+d_{\nu,\text{DMA}})\\
&\hspace{1.5cm}\times P_1^{(q)}(i,\sigma)P_1{(q)}(j,\sigma)\Big].
\end{split}\end{equation}

\subsection{$\boldsymbol{f_\nu^2(s)}$ of DFA in the increment representation}\label{sec:flucfunwedfa}
For DFA the estimator of the generalised squared path displacement is
\begin{equation}\begin{split}
&f^2_{\nu,\text{DFA$q$}}(s)\\
&=\frac{1}{s} \sum_{t=1}^s\Big(y(t+d_{\nu,\text{DFA}})-p_\nu^{(q)}(t+d_{\nu,\text{DFA}})\Big)^2,
\end{split}\end{equation}
see Eq.~(\ref{vardfa}). Using the residual as weighted sum given in Eq.~(\ref{resweight}) we can write this as
\begin{equation}\begin{split}\label{ret2}
&f_{\nu,\text{DFA}q}^2(s) \\
&= \frac{1}{s} \sum_{t=1}^s \left( \sum_{i=1}^s x(i+d_{\nu,\text{DFA}}) \omega^{(q)}(i,t) \right)^2, \\
&= \sum_{i,j=1}^s x(i+d_{\nu,\text{DFA}})x(j+d_{\nu,\text{DFA}}) \\
&\hspace{1.5cm}\times\frac{1}{s} \sum_{t=1}^s \omega^{(q)}(i,t)  \omega^{(q)}(j,t) .
\end{split}\end{equation}
Hence the weights of the DFA fluctuation function are
\begin{equation}
L_{\text{DFA}q}(i,j,s) = \frac{1}{s} \sum_{t=1}^s \omega^{(q)}(i,t)  \omega^{(q)}(j,t)
\end{equation}
which has no breaks in contrast to the weights of DMA. In detail for $j>i$ it can be written as
\begin{equation}\begin{split}\label{dfaweight65}
&L_{\text{DFA}q}(i,j,s) \\
&= \frac{1}{s} \Bigg( \sum_{t=j}^s 1 - \sum_{t=i}^sP-1^{(q)}(j,t)\\
&- \sum_{t=j}^sP_1^{(q)}(i,t)+\sum_{t=1}^s P_1^{(q)}(i,t)P_1^{(q)}(j,t)\Bigg).
\end{split}\end{equation}
The restriction $j>i$ is no problem because the double sum $\sum_{i,j=1}^s$ can ordered such that only terms with $j>i$ occur, see Eq.~(\ref{nsordering}).

\section{Stationary DMA fluctuation function}\label{sec:statdmafluc}
The fluctuation function of DMA for stationary processes is given by
 \begin{equation}\begin{split}\label{flucheredma}
F_{\text{DMA}q}^2(s)&=\langle x^2(i)\rangle \Big( \mathcal{L}_{\text{DMA}q}(0,s)\\
&+ 2 \sum_{\tau=1}^{s-1} C(\tau)  \mathcal{L}_{\text{DMA}q}(\tau,s)\Big),
 \end{split}\end{equation}
see Eq.~(\ref{flucstat222}) which can be written more detailed as follows. The weights $\mathcal{L}_{\text{DMA}q}(\tau,s)$ have different expressions depending on the time lag $\tau$. This can be derived by using $L_{\text{DMA}q}(i,j,s)$ of Eq.~(\ref{gdfz}) in $\mathcal{L}_{\text{DMA}q}(\tau,s) = \sum_{i=1}^{s-\tau} L_{\text{DMA}q}(i,i+\tau,s)$ of Eq.~(\ref{statweight1}). Again $\sigma=(s+1)/2$. Hence we find the following. For $1\le\tau \le \sigma-2$ it is
\begin{equation}\begin{split}\label{w1}
&\mathcal{L}_{\text{DMA}q}(\tau,s)\\
 &= \sum_{i=1}^{\sigma-\tau}\left(1-P_1^{(q)}(i,\sigma)\right)\left(1-P_1^{(q)}(i+\tau,\sigma)\right) \\
&+  \sum_{i=\sigma-\tau+1}^{\sigma}\left(1-P_1^{(q)}(i,\sigma)\right)\left(-P_1^{(q)}(i+\tau,\sigma)\right)\\
&+ \sum_{i=\sigma+1}^{s-\tau}P_1^{(q)}(i,\sigma)P_1^{(q)}(i+\tau,\sigma).
\end{split}\end{equation}
For $\tau = \sigma-1$ it is
\begin{equation}\begin{split}\label{w2}
&\mathcal{L}_{\text{DMA}q}(\tau,s) \\
&= \sum_{i=1}^{\sigma-\tau}\left(1-P_1^{(q)}(i,\sigma)\right)\left(1-P_1^{(q)}(i+\tau,\sigma)\right) \\
&+  \sum_{i=\sigma-\tau+1}^{s-\tau}\left(1-P_1^{(q)}(i,\sigma)\right)\left(-P_1^{(q)}(i+\tau,\sigma)\right).
\end{split}\end{equation}
For $\sigma \le \tau \le s-1$ it is
\begin{equation}\begin{split}\label{w3}
&\mathcal{L}_{\text{DMA}q}(\tau,s) \\
&= \sum_{i=1}^{s-\tau}\left(1-P_1^{(q)}(i,\sigma)\right)\left(-P_1^{(q)}(i+\tau,\sigma)\right).
\end{split}\end{equation}
Let us write the weights of Eq.~(\ref{w1}), (\ref{w2}) and (\ref{w3}) as $\mathcal{L}_{\text{DMA}q}^{(1)}(\tau,s)$, $\mathcal{L}_{\text{DMA}q}^{(2)}(\tau,s)$ and $\mathcal{L}_{\text{DMA}q}^{(3)}(\tau,s)$. Then the statioary fluctuation function of Eq.~(\ref{flucheredma}) is given in detail by
\begin{equation}\begin{split}\label{statflucdmafull}
&F_{\text{DMA}q}^2(s) \\
&= \langle x^2(i)\rangle \Bigg(\mathcal{L}_{\text{DMA}q}^{(1)}
(0,s)+2\sum_{\tau=1}^{\sigma-2}C(\tau)\mathcal{L}_{\text{DMA}q}^{(1)}(\tau,s)\\
 &+2\sum_{\tau=\sigma-1}^{\sigma-1}C(\tau)\mathcal{L}_{\text{DMA}q}^{(2)}(\tau,s)+2\sum_{\tau=\sigma}^{s-1}C(\tau)\mathcal{L}_{\text{DMA}q}^{(3)}(\tau,s)\Bigg).
\end{split}\end{equation}

\section{Autocovoariance difference of FBM}\label{sec:fbmdfiff}
The autocovariance difference of FBM is
\begin{equation}
D_\nu(i,j) = \frac{1}{2} \left((i+d_\nu)^{2H}+(j+d_\nu)^{2H}-i^{2H}-j^{2H} \right),
\end{equation}
see Eq.~(\ref{acovdifffbm}). We can write $(i+d_\nu)^{2H}-i^{2H}$ as infinite sum
\begin{equation}
(i+d_\nu)^{2H}-i^{2H} = \sum_{m=0}^\infty i^m (v_m(h,d_\nu,r)-v_m(h,0,r))
\end{equation}
using Taylor's series, see appendix \ref{sec:fbmdfiff}.\ref{appendix:taylor} for the definition of $v_m$. The same holds for the $j$-dependent term of $D_\nu(i,j)$. Hence we can write the autocovariance difference as
\begin{equation}
D_\nu(i,j) = \sum_{m=0}^\infty (i^m+j^m) (v_m(h,d_\nu,r)-v_m(h,0,r)).
\end{equation}
Note that $r$ can be chosen arbitrarily. In section \ref{sec:biasint} we use the definition $\lambda_m=v_m(h,d_\nu,r)-v_m(h,0,r)$.
\subsection{Taylor series}\label{appendix:taylor}
For the function
\begin{equation}
f(i) = (i+d)^h
\end{equation}
with constants $d\ge 0$ and $h>0$ the Taylor series at point $r$ is
\begin{equation}
f(i) = \sum_{n=0}^\infty \frac{f^{(n)}(i_0)}{n!}(i-r)^n.
\end{equation}
With the derivative
\begin{equation}
f^{(n)}(i) = \begin{pmatrix}h\\n\end{pmatrix} n! (i+d)^{h-n}.
\end{equation}
the function is
\begin{equation}
f(i) = \sum_{n=0}^\infty \begin{pmatrix}h\\n\end{pmatrix} (r+d)^{h-n}(i-r)^n.
\end{equation}
Using binomial theorem for the term
\begin{equation}
(i-r)^n = \sum_{m=0}^n \begin{pmatrix}n\\m\end{pmatrix} (-r)^{n-m}i^n.
\end{equation}
the function is
\begin{equation}
f(i) = \sum_{n=0}^\infty\sum_{m=0}^n \begin{pmatrix}h\\n\end{pmatrix}\begin{pmatrix}n\\m\end{pmatrix} (r+d)^{h-n}  (-r)^{n-m}i^m.
\end{equation}
This can be ordered with respect to $i$, namely
\begin{equation}
f(i) = \sum_{m=0}^\infty i^m v_m(h,d,r)
\end{equation}
with
\begin{equation}
v_m(h,d,r) = \sum_{n=m}^\infty \begin{pmatrix}h\\n\end{pmatrix}\begin{pmatrix}n\\m\end{pmatrix} (r+d)^{h-n}  (-r)^{n-m}.
\end{equation}
If $d=0$ we pick a nonzero point $r \ne 0$.

\section{Mathematica codes}\label{sec:mathematica}
Here we provide Mathematica codes for important quantitites of this article. They can easily be used by simply copying the source code into Mathematica and execute them. We indicate the begin of a line in the code with (**) for the sake of clarity. Furthermore we explain how the codes can be modified to get different outputs.
\subsection{Weights of the first fit}\label{sec:appwei}
Here we provide the Mathematica code for the weights of the fit with $\nu=1$, namely
\begin{equation}
P_1^{(q)}(i,t)=\sum_{m=0}^q t^m \sum_{n=0}^q  \left(\mathbb{S}_1^{-1}\right)_{m+1,n+1} \sum_{k=i}^s k^n,
\end{equation}
see Eq.~(\ref{fitweight}). This is calculated by the following Mathematica code:
\begin{verbatim}
(**)q=0;
(**)S[i_]:=Sum[k^i,{k,1,s}];
(**)matrixS:=Table[S[m+n-2],{m,q+1},{n,q+1}];
(**)Simplify[Sum[t^m*Sum[Inverse[matrixS][[
    m+1,n+1]]*Sum[k^n,{k,i,s}],{n,0,q}],{
    m,0,q}]]
    \end{verbatim}
The output of this code gives $P_1^{(0)}(i,t)$ with order of detrending $q=0$. The order of detrending $q$ in the first line "\verb|q=0;|" can be changed to any order, e.g. to "\verb|q=1;|" for first order of detrending. Note that $P_1^{(q)}(i,t)$ is the same for DFA and DMA.

\subsection{Weights of the fits}\label{sec:appwei2}
Here we provide the Mathematica code for the weights of the fit for DFA and DMA
\begin{equation}
P_\nu^{(q)}(i,t)= \sum_{m=0}^q (t+d_\nu)^m \sum_{n=0}^q (\mathbb{S}_\nu^{-1})_{m+1,n+1} \sum_{k=i+d_\nu}^{s+d_\nu} k^n,
\end{equation}
see Eq.~(\ref{fitweight}). This is calculated by the following Mathematica code:
\begin{verbatim}
(**)q=0;
(**)d[v_]:=(v-1)*s;
(**)S[i_]:=Sum[k^i,{k,1+d[v],s+d[v]}];
(**)matrixS:=Table[S[m+n-2],{m,q+1},{n,q+1}];
(**)Simplify[Sum[(t+d[v])^m*Sum[Inverse[
    matrixS][[m+1,n+1]]*Sum[k^n,{k,i+d[v],s+
    d[v]}],{n,0,q}],{m,0,q}]]
\end{verbatim}
The output of this code gives $P_{\nu,\text{DFA}}^{(0)}(i,t)$ with order of detrending $q=0$ and DFA shift factor $d_{\nu,\text{DFA}}=(\nu-1)s$. The order of detrending in first line "\verb|q=0;|" can be changed to any order, e.g. to "\verb|q=1;|" for first order of detrending. The DFA shift factor in the second line "\verb|d[v_]:=(v-1)*s;|" can be changed to the DMA shift factor "\verb|d[v_]:=v-1;|". Note that $P_\nu^{(q)}(i,t)$ is the same for DFA and DMA.

\subsection{Weights of DFA fluctuation function}\label{weightsflucmat}
Here we provide the Mathematica code for the weights of the DFA fluctuation function with $j> i$, namely
\begin{equation}\begin{split}
&L_{\text{DFA}q}(i,j,s) \\
&= \frac{1}{s} \Bigg( \sum_{t=j}^s 1 - \sum_{t=i}^sP_1^{(q)}(j,t)\\
&- \sum_{t=j}^sP_1^{(q)}(i,t)+\sum_{t=1}^s P_1^{(q)}(i,t)P_1^{(q)}(j,t)\Bigg),
\end{split}\end{equation}
see Eq.~(\ref{dfaweight65}). This is calculated by the following Mathematica code:
\begin{verbatim}
(**)P[i_,t_]:=(1-i+s)/s;
(**)Simplify[1/s*(Sum[1,{t,j,s}]-Sum[P[j,t],
    {t,i,s}]-Sum[P[i,t],{t,j,s}]+Sum[P[i,t]*
    P[j,t],{t,1,s}])]
\end{verbatim}
The output of this code gives $L_{\text{DFA}0}(i,j,s)$ using the output from appendix \ref{sec:mathematica}.\ref{sec:appwei} which is the weight of the fit $P_1^{(0)}(i,t)=(1-i+s)/s$. This output from appendix \ref{sec:mathematica}.\ref{sec:appwei} in the first line "\verb|(1-i+s)/s|" can be changed to any output from the code of appendix \ref{sec:mathematica}.\ref{sec:appwei} by changing the definition of "\verb|q|" in the code of appendix \ref{sec:mathematica}.\ref{sec:appwei} as explained there.

\subsection{$\boldsymbol{\langle f_\nu^2(s)\rangle}$ of DMA}\label{fv2dma}
Here we provide the Mathematica code for the mean of the estimator of the generalised squared path displacement of DMA in the increment representation
\begin{equation}
\langle f^2_{\nu,\text{DMA}q}(s) \rangle = \sum_{i,j=1}^s G_\nu(i,j)L_{\text{DMA}q}(i,j,s),
\end{equation}
see Eq.~(\ref{ret3}). The function $G_\nu(i,j)$ can in principle be arbitrary. Here in the article it is $G_\nu(i,j)=\langle x(i+d_{\nu,\text {DMA}})x(j+d_{\nu,\text {DMA}})\rangle$, $\text{Cov}(i,j)$ and $D_\nu(i,j)$. But due to the possible complicated form of $G_\nu(i,j)$ the code will not always be succesfully calculating a result. With the help of Eq.~(\ref{fdmadetail}) it is now in detail
\begin{equation}\begin{split}
&\langle f^2_{\nu,\text{DMA}q}(s) \rangle  \\
&= \sum_{i=1}^\sigma  \Big[ G_\nu(i,i)\left(1-P_1^{(q)}(i,\sigma)\right)^2 \Big] \\
&+ \sum_{i=\sigma+1}^s \Big[ G_\nu(i,i)\left(P_1^{(q)}(i,\sigma)\right)^2 \Big]\\
&+2\sum_{i=1}^\sigma\sum_{j=i+1}^\sigma \Big[ G_\nu(i,j)\\
&\hspace{1.5cm}\times\left(1-P_1^{(q)}(i,\sigma)\right)\left(1-P_1^{(q)}(j,\sigma)\right)\Big]\\
&+2\sum_{i=1}^\sigma\sum_{j=\sigma+1}^s \Big[G_\nu(i,j)\\
&\hspace{1.5cm}\times\left(1-P_1^{(q)}(i,\sigma)\right)\left(-P_1^{(q)}(j,\sigma)\right) \Big]\\
&+2\sum_{i=\sigma+1}^{s-1} \sum_{j=i+1}^s \Big[ G_\nu(i,j)P_1^{(q)}(i,\sigma)P_1^{(q)}(j,\sigma)\Big].
\end{split}\end{equation}
This is calculated by the following Mathematica code:
\begin{verbatim}
(**)P[i_,t_]:=(1-i+s)/s;
(**)d[v_]=v-1;
(**)p=1;
(**)G[i_,j_,v_]:=(i+d[v])^p*(j+d[v])^p;
(**)sigma=(s+1)/2;
(**)Simplify[Sum[G[i,i,v]*(1-P[i,sigma])^2,
    {i,1,sigma}]+Sum[G[i,i,v]*P[i,sigma]^2,
    {i,sigma+1,s}]+2*Sum[Sum[G[i,j,v]*(1-
    P[i,sigma])*(1-P[j,sigma]),{j,i+1,
    sigma}],{i,1,sigma}]+2*Sum[Sum[G[i,j,
    v]*(1-P[i,sigma])*(-P[j,sigma]),{j,
    sigma+1,s}],{i,1,sigma}]
    +2*Sum[Sum[G[i,j,v]*P[i,sigma]*P[j,
    sigma],{j,i+1,s}],{i,sigma+1,s-1}]]
\end{verbatim}
The output of this code gives $\langle f^2_{\nu,\text{DMA}0}(s) \rangle$ using the output from appendix \ref{sec:mathematica}.\ref{sec:appwei} which is the weight of the fit $P_1^{(0)}(i,t)=(1-i+s)/s$ and the autocovariance difference of an additive trend $G_{ij}^{(\nu)}=(i+d_{\nu,\text{DMA}})^p(i+d_{\nu,\text{DMA}})^p$ of a trend with order $p=1$. The output from appendix \ref{sec:mathematica}.\ref{sec:appwei} in the first line "\verb|(1-i+s)/s|" can be changed to any output from the code of appendix \ref{sec:mathematica}.\ref{sec:appwei} by changing the definition of "\verb|q|" in the code of appendix \ref{sec:mathematica}.\ref{sec:appwei} as explained there. The autocovariance difference of the trend in the fourth line "\verb|(i+d[v])^p*(j+d[v])^p|" can be changed to any $G_\nu(i,j)$, e.g. the autocovariance difference of FBM "\verb|i^p+j^p|" where we left out the prefactors. The order of $p$ in the third line "\verb|p=1|" can be changed to any order, e.q. to "\verb|p=2|" for second order.

\subsection{$\boldsymbol{\langle f_\nu^2(s)\rangle}$ of DFA}\label{fv2dfa}
Here we provide the Mathematica code for the mean of the estimator of the generalised squared path displacement of DMA in the increment representation
\begin{equation}\label{gddsdff}
\langle f^2_{\nu,\text{DFA}q}(s) \rangle = \sum_{i,j=1}^s G_\nu(i,j)L_{\text{DFA}q}(i,j,s),
\end{equation}
see Eq.~(\ref{ret2}). The function $G_\nu(i,j)$ can in principle be arbitrary. Here in the article it is $G_\nu(i,j)=\langle x(i+d_{\nu,\text {DFA}})x(j+d_{\nu,\text {DFA}})\rangle$, $\text{Cov}(i,j)$ and $D_\nu(i,j)$. But due to the possible complicated form the code will not always be succesfully calculating a result. We can order Eq.~(\ref{gddsdff}) to
\begin{equation}\begin{split}\label{nsordering}
\langle f^2_{\nu,\text{DFA}q}(s) \rangle &=  \sum_{i=1}^s G_\nu(i,i)L_{\text{DFA}q}(i,i,s)\\
&+2\sum_{i=1}^{s-1}\sum_{j=i+1}^s G_\nu(i,j)
L_{\text{DFA}q}(i,j,s)
\end{split}\end{equation}
This is calculated by the following Mathematica code:
\begin{verbatim}
(**)L[i_,j_,s_]:=((-1+i) (1-j+s))/s^2;
(**)d[v_]=(v-1)*s;
(**)p=1;
(**)G[i_,j_,v_]:=(i+d[v])^p*(j+d[v])^p;
(**)Simplify[Sum[G[i,i,v]*L[i,i,s],{i,1,s}]
    +2*Sum[Sum[G[i,j,v]*L[i,j,s],{j,i+1,s}],
    {i,1,s-1}]]
\end{verbatim}
The output of this code gives $\langle f^2_{\nu,\text{DFA}0}(s) \rangle $ using the output from appendix \ref{sec:mathematica}.\ref{weightsflucmat} which is the weight of the fluctuation function $L_{\text{DFA}0}(i,j,s)=(i-1)(s-j+1)/s^2$ and the autocovariance difference of an additive trend $G_{ij}^{(\nu)}=(i+d_{\nu,\text{DMA}})^p(i+d_{\nu,\text{DMA}})^p$ of a trend with order $p=1$. The output from appendix \ref{sec:mathematica}.\ref{weightsflucmat} in the first line "\verb|((-1+i) (1-j+s))/s^2|" can be changed to any output from the code of appendix \ref{sec:mathematica}.\ref{weightsflucmat} by changing the definition of "\verb|P[i_,t_]|" in the code of appendix \ref{sec:mathematica}.\ref{weightsflucmat} as explained there. The autocovariance difference of the trend in the fourth line "\verb|(i+d[v])^p*(j+d[v])^p|" can be changed to any $G_\nu(i,j)$, e.g. the autocovariance difference of FBM "\verb|i^p+j^p|" where we left out the prefactors. The order of $p$ in the third line "\verb|p=1|" can be changed to any order, e.q. to "\verb|p=2|" for second order.

\subsection{$\boldsymbol{\langle \widehat{F^2}(s) \rangle }$ of DMA and DFA}\label{meanestfluc}
Here we provide the Mathematica code for the mean of the estimator of the fluctuation function of DMA and DFA
\begin{equation}
\langle \widehat{F^2}(s) \rangle = \frac{1}{K} \sum_{\nu=1}^K \langle f^2_\nu(s) \rangle,
\end{equation}
see Eq.~(\ref{flucpath}). This is calculated by the following code:
\begin{verbatim}
(**)f[v_,s_]:=1/180(-1+s^2)(14+15s(-
    1+2v)+s^2(4-15v+15 v^2));
(**)k=n/s;
(**)Simplify[1/k*Sum[f[v,s],{v,1,k}]]
\end{verbatim}
The output of this code gives $\langle \widehat{F^2}_{\text{DFA}0}(s) \rangle$ using $K=N/s$ and the output from appendix \ref{sec:mathematica}.\ref{fv2dfa} which is the mean of the estimator of the generalised squared path displacement of DFA $\langle f^2_{\nu,\text{DFA}0}(s) \rangle $. This output from appendix \ref{sec:mathematica}.\ref{fv2dfa} in the first line "\verb|1/180(-1+s^2...|" can be changed to any output from the code of appendix \ref{sec:mathematica}.\ref{fv2dfa} by changing the definition of "\verb|L[i_,j_,s_]|" in the code of appendix \ref{sec:mathematica}.\ref{fv2dfa} as explained there. Furthermore can the first line "\verb|1/180(-1+s^2...|" be changed to the output from the code of appendix \ref{sec:mathematica}.\ref{fv2dma} which is the mean of the estimator of the generalised squared path displacement of DMA $\langle f^2_{\nu,\text{DMA}0}(s) \rangle $. This output from appendix \ref{sec:mathematica}.\ref{fv2dma} can be changed to any output from the code of appendix \ref{sec:mathematica}.\ref{fv2dma} by changing the definition of "\verb|P[i_,t_]|" in the code of appendix \ref{sec:mathematica}.\ref{fv2dma} as explained there. For DMA the second line "\verb|k=n/s|" has to be changed to "\verb|k=n-s+1|".\\ \\

\subsection{Weights of stationary DMA fluctuation function}\label{sec:dmastatweight}
Here we provide the Mathematica code for the weights of the stationary fluctuation function of DMA $\mathcal{L}_{\text{DMA}q}^{(1)}(\tau,s)$, $\mathcal{L}_{\text{DMA}q}^{(2)}(\tau,s)$ and $\mathcal{L}_{\text{DMA}q}^{(3)}(\tau,s)$, see for their explicit forms Eq.~(\ref{w1}), (\ref{w2}) and (\ref{w3}) in appendix \ref{sec:statdmafluc}. These are calculated by the following code:
\begin{verbatim}
(**)P[i_,t_]:=(1-i+s)/s;
(**)sigma=(s+1)/2;
(**)Simplify[Sum[(1-P[i,sigma])*(1-P[i
    +tau,sigma]),{i,1,sigma-tau}]+Sum[
    (1-P[i,sigma])*(-P[i+tau,sigma]),
    {i,sigma-tau+1,sigma}]+Sum[P[i,
    sigma]*P[i+tau,sigma],{i,sigma+1,
    s-tau}]]
(**)Simplify[Sum[(1-P[i,sigma])*(1-P[i
    +tau,sigma]),{i,1,sigma-tau}]+Sum[
    (1-P[i,sigma])*(-P[i+tau,sigma]),
    {i,sigma-tau+1,s-tau}]]
(**)Simplify[Sum[(1-P[i,sigma])*(-P[i+
    tau,sigma]),{i,1,s-tau}]]
\end{verbatim}
The output of this code gives  $\mathcal{L}_{\text{DMA}0}^{(1)}(\tau,s)$, $\mathcal{L}_{\text{DMA}0}^{(2)}(\tau,s)$ and $\mathcal{L}_{\text{DMA}0}^{(3)}(\tau,s)$ using the output from appendix \ref{sec:mathematica}.\ref{sec:appwei} which is the weight of the fit $P_1^{(0)}(i,t)=(1-i+s)/s$.  This output from appendix \ref{sec:mathematica}.\ref{sec:appwei} in the first line "\verb|(1-i+s)/s|" can be changed to any output from the code of appendix \ref{sec:mathematica}.\ref{sec:appwei} by changing the definition of "\verb|q|" in the code of appendix \ref{sec:mathematica}.\ref{sec:appwei} as explained there.

\subsection{Weights of stationary DFA fluctuation function}\label{sec:dfastatweight}
Here we provide the Mathematica code for the weights of the stationary fluctuation function of DFA
 \begin{equation}
 \mathcal{L}_{\text{DFA}q}(\tau,s) =  \sum_{i=1}^{s-\tau} L_{\text{DFA}q}(i,i+\tau,s),
  \end{equation}
see Eq.~(\ref{statweight1}). This is calculated by the following code:
\begin{verbatim}
(**)L[i_,j_,s_]:=((-1+i) (1-j+s))/s^2;
(**)Simplify[Sum[L[i,i+tau,s],{i,1,s-tau}]]
\end{verbatim}
The output of this code gives $ \mathcal{L}_{\text{DFA}0}(\tau,s)$ using the output from appendix \ref{sec:mathematica}.\ref{weightsflucmat} which is the weight of the fluctuation function $L_{\text{DFA}0}(i,j,s)=(i-1)(s-j+1)/s^2$. This output from appendix \ref{sec:mathematica}.\ref{weightsflucmat} in the first line "\verb|((-1+i) (1-j+s))/s^2|" can be changed to any output from the code of appendix \ref{sec:mathematica}.\ref{weightsflucmat} by changing the definition of "\verb|P[i_,t_]|" in the code of appendix \ref{sec:mathematica}.\ref{weightsflucmat} as explained there.

\subsection{Stationary DMA fluctuation function}\label{sec:dmaflucmath}
Here we provide the Mathematica code for the stationary fluctuation function of DMA
\begin{equation}\begin{split}
&F_{\text{DMA}q}^2(s) \\
&= \langle x^2(i)\rangle \Bigg(\mathcal{L}_{\text{DMA}q}^{(1)}
(0,s)+2\sum_{\tau=1}^{\sigma-2}C(\tau)\mathcal{L}_{\text{DMA}q}^{(1)}(\tau,s)\\
 &+2\sum_{\tau=\sigma-1}^{\sigma-1}C(\tau)\mathcal{L}_{\text{DMA}q}^{(2)}(\tau,s)+2\sum_{\tau=\sigma}^{s-1}C(\tau)\mathcal{L}_{\text{DMA}q}^{(3)}(\tau,s)\Bigg).
\end{split}\end{equation}
see Eq.~(\ref{statflucdmafull}). This is calculated by the following code:
\begin{verbatim}
(**)c[tau_]:=a^tau/(1-a^2);
(**)sigma=(s+1)/2;
(**)L1[tau_]:=(s^3-6 s^2 tau+2 tau (-1+
    tau^2)+s (-1+6 tau^2))/(12 s^2);
(**)L2[tau_]:=(s-s^3-4 tau+4 tau^3)/
    (24 s^2);
(**)L3[tau_]:=-(((1+s-tau) (-s+s^2+tau-
    2 s tau+tau^2))/(6 s^2));
(**)Simplify[c[0]*L1[0]+2*Sum[c[tau]*
    L1[tau],{tau,1,sigma-2}]+2*Sum[c[
    tau]*L2[tau],{tau,sigma-1,sigma-1}]
    +2*Sum[c[tau]*L3[tau],{tau,sigma,s-1}]]
\end{verbatim}
The output of this code gives $F_{\text{DMA}0}^2(s)$ using the autocorrelation function of an AR($1$) process $\text{Cov}(\tau)=a^\tau/(1-a^2)$ with parameter $a$ and the outputs of appendix \ref{sec:mathematica}.\ref{sec:dmastatweight} which are the weights of the stationary fluctuation function stationary of DMA $\mathcal{L}_{\text{DMA}0}^{(1)}(\tau,s)$, $\mathcal{L}_{\text{DMA}0}^{(2)}(\tau,s)$ and $\mathcal{L}_{\text{DMA}0}^{(3)}(\tau,s)$. The autocovariance function in the first line "\verb|c[tau_]:=a^tau/(1-a^2);|" can be changed to any autocovariance function, e.g. the autocovariance function of white noise with unit variance "\verb|c[tau_]:=KroneckerDelta[tau,0];|" or ARFIMA($0$,$d$,$0$). The autocovariance function of an ARFIMA($0$,$d$,$0$) is $\text{Cov}(\tau)=\Gamma(\tau+d)\Gamma(\tau-2d)/(\Gamma(\tau+1-d)\Gamma(1-d)\Gamma(d))$.  The outputs from appendix \ref{sec:mathematica}.\ref{sec:dmastatweight} in the third line "\verb|(s^3-6s^2tau+...|", fourth line "\verb|(s-s^3-...|" and fifth line "\verb|(((1+s-tau)...|" can be changed to any outputs from appendix \ref{sec:mathematica}.\ref{sec:dmastatweight} by changing the definition of "\verb|P[i_,t_]|" in the code from appendix \ref{sec:mathematica}.\ref{sec:dmastatweight} as explained there.

\subsection{Stationary DFA fluctuation function}\label{sec:dfaflucmath}
Here we provide the Mathematica code of the stationary fluctuation function of DFA
 \begin{equation}\begin{split}
F_{\text{DFA}q}^2(s)&=\langle x^2(i)\rangle \Big( \mathcal{L}_{\text{DFA}q}(0,s)\\
&+ 2 \sum_{\tau=1}^{s-1} C(\tau)  \mathcal{L}_{\text{DFA}q}(\tau,s)\Big),
 \end{split}\end{equation}
see Eq.~(\ref{flucstat222}). This is calculated by the following code:
\begin{verbatim}
(**)c[tau_]:=a^tau/(1-a^2);
(**)L[tau_]:=((1+s-tau)(-s+s^2+tau-
    2 s tau+tau^2))/(6 s^2);
(**)Simplify[c[0]*L[0]+2*Sum[c[
    tau]*L[tau],{tau,1,s-1}]]
\end{verbatim}
The output of this code gives $F_{\text{DFA}0}^2(s)$ using the autocorrelation function of an AR($1$) process $\text{Cov}(\tau)=a^\tau/(1-a^2)$ with parameter $a$ and the output of appendix \ref{sec:mathematica}.\ref{sec:dfastatweight} which are the weights of the stationary fluctuation function of DFA $\mathcal{L}_{\text{DFA}0}(\tau,s$.) The autocovariance function in the first line "\verb|c[tau_]:=a^tau/(1-a^2);|" can be changed to any autocovariance function, e.g. the autocovariance function of white noise with unit variance "\verb|c[tau_]:=KroneckerDelta[tau,0];|" or ARFIMA($0$,$d$,$0$). The autocovariance function of an ARFIMA($0$,$d$,$0$) is $\text{Cov}(\tau)=\Gamma(\tau+d)\Gamma(\tau-2d)/(\Gamma(\tau+1-d)\Gamma(1-d)\Gamma(d))$.  The output from appendix \ref{sec:mathematica}.\ref{sec:dfastatweight} in the second line "\verb|((1+s-tau)...|" can be changed to any output from appendix \ref{sec:mathematica}.\ref{sec:dfastatweight} by changing the definition of "\verb|L[i_,j_,s]|" in the code from appendix \ref{sec:mathematica}.\ref{sec:dfastatweight} as explained there.

\end{appendix}


\begin{thebibliography}{sotief}

\bibitem{hurst} H.E. Hurst, T. Am. Soc. Civ. Eng. \textbf{116}, 770 (1951).

\bibitem{graves} T. Graves, R. Gramacy, N. Watkins and C. Franzke, Entropy \textbf{19}, 437 (2017).
    
\bibitem{palma} W. Palma, \textit{Long-Memory Time Series}, Wiley, 2007

\bibitem{beran} J. Beran, \textit{Statistics for Long-Memory Processes}, Chapmann \& Hall, 1994

\bibitem{pipiras} V. Pipiras and M.S. Taqqu, \textit{Long-Range Dependence and Self-Similarity}, Cambridge University Press, 2017
    
\bibitem{kantelhardt2} J.W. Kantelhardt, E. Koscielny-Bunde, H.H.A. Rego, S. Havlin and A. Bunde, Physica A \textbf{295}, 441 (2001).
    
\bibitem{peng} C.K. Peng, S.V. Buldyrev, S. Havlin, M. Simons, H.E. Stanley and A.L. Goldberger, Phys. Rev. E \textbf{49}, 1685 (1994).
    
\bibitem{penzel} T. Penzel, J.W. Kantelhardt, L. Grote, J.H. Peter and A. Bunde, IEEE T. Bio-Med. Eng. \textbf{50}, 1143 (2003).

\bibitem{echeverria} J.C. Echeverria, M.S. Woolfson, J.A. Crowe, B.R. Hayes-Gill, G.D.H. Croaker and H. Vyas, Chaos \textbf{13}, 467 (2003).

\bibitem{castiglioni} P. Castiglioni, G. Parati, A. Civijian, L. Quintin, and M. Di Rienzo, IEEE T. Bio-Med. Eng. \textbf{56}, 675 (2009).

\bibitem{baumert} M. Baumert, M. Javorka, A. Seeck, R. Faber, P. Sanders, and A. Voss, Comput. Biol. Med. \textbf{42}, 347 (2012).

\bibitem{talkner} P. Talkner and R.O. Weber, Phys. Rev. E \textbf{62}, 150 (2000).

\bibitem{kiraly} A. Kiraly and I.M. Janosi, Meteorol. Atmos.
Phys. \textbf{88}, 119 (2005).

\bibitem{kurnaz}  M.L. Kurnaz, Fractals \textbf{12}, 365 (2004).

\bibitem{bunde} E. Koscielny-Bunde, A. Bunde, S. Havlin, H.E. Roman, Y. Goldreich and H.J. Schellnhuber, Phys. Rev. Lett. \textbf{81}, 729 (1998).

\bibitem{meyer} P. Meyer, M. Hoell and H. Kantz, J. Geophys. Res. Atmos. \textbf{123}, 4413 (2018).

\bibitem{massah} M. Massah and H. Kantz, Geophys. Res. Lett. \textbf{43}, 9243 (2016).

\bibitem{zhang} Q. Zhang, Y. Zhou and V.P. Singh, Hydrol. Process. \textbf{28}, 753 (2014).

\bibitem{zhang3} Q. Zhang, Y. Zhou, V.P. Singh and X. Chen, Hydrol. Process. \textbf{26}, 436 (2012). 

\bibitem{ivanova} K. Ivanova and M. Ausloos, Physica A \textbf{274}, 349 (1999).

\bibitem{luo} M. Luo, Y. Leung, Y. Zhou and W. Zhang, J. Clim. \textbf{28}, 3122 (2015).

\bibitem{cao} G. Cao and M. Guangxi, Physica A \textbf{436}, 25 (2015).

\bibitem{reboredo} J.C Reboredo, M.A. Rivera-Castro, J.G.V. Miranda and R. Garcia-Rubio, Physica A \textbf{392}, 1631 (2013).

\bibitem{serletis} A. Serletis, O.Y. Uritskaya and V.M. Uritsky, Int. J. Bifurcat. Chaos \textbf{18}, 599 (2008).

\bibitem{ramirez} J. Alvarez-Ramirez, J. Alvarez and E. Rodriguez, Energ. Econ. \textbf{30}, 2645 (2008).

\bibitem{wang} Y. Wang, Y. Wei and C. Wu, Physica A \textbf{390}, 864 (2011).

\bibitem{taqqu2} M.S. Taqqu, V. Teverovsky and W. Wilinger, Fractals \textbf{3}, 785 (1995).
        
\bibitem{hu} K. Hu, P.C. Ivanov, Z. Chen, P. Carpena and H.E. Stanley, Phys. Rev. E \textbf{64}, 011114 (2001).

\bibitem{chen} Z. Chen, P.C. Ivanov, K. Hu and H.E. Stanley, Phys. Rev. E \textbf{65}, 041107 (2002). 
    
\bibitem{chen2} Z. Chen, K. Hu, P. Carpena, P. Bernaola-Galvan, H.E. Stanley and P.C. Ivanov, Phys. Rev. E \textbf{71}, 011104, (2005).

\bibitem{xu} L.M. Xu, P.C. Ivanov, K. Hu, Z. Chen, A. Carbone and H.E. Stanley, Phys. Rev. E \textbf{71}, 051101 (2005).

\bibitem{bashan} A. Bashan, R. Bartsch, J.W. Kantelhardt and S. Havlin, Physica A \textbf{387}, 5080 (2008).
    
\bibitem{ma} Q.D.Y. Ma, R.P. Bartsch, P. Bernaola-Galvan, M. Yoneyama and P.C. Ivanov, Phys. Rev. E \textbf{81}, 031101 (2010). 

\bibitem{weron} R. Weron, Physica A \textbf{312}, 285 (2002).

\bibitem{kantelhardt} J.W. Kantelhardt, S.A. Zschiegner, E. Koscielny-Bunde, S. Havlin, A. Bunde and H.E. Stanley, Physica A \textbf{316}, 87 (2002).

\bibitem{movahed} M.S. Movahed, G.R. Jafari, F. Ghasemi, S. Rahvar and M.R.R. Tabar, J. Stat. Mech.-Theory E., P02003, (2006).

\bibitem{movahed2} M.S. Movahed and E. Hermanis, Physica A \textbf{387}, 915 (2008).

\bibitem{zhang2} Q. Zhang, Y. Zhou, V.P. Singh and Y.D. Chen, J. Hydrol. \textbf{400}, 121 (2011).

\bibitem{zhou} Y. Zhou, Y. Leung, and Z.G. Yu, Phys. Rev. E \textbf{87}, 012921 (2013).
 
\bibitem{zhou2} Y. Zhou and Y. Leung, J. Stat. Mech.-Theory E., P06021 (2010).

\bibitem{podobnik} B. Podobnik and H.E. Stanley, Phys. Rev. Lett. \textbf{100}, 084102 (2008).

\bibitem{zhou3} W.X. Zhou, Phys. Rev. E \textbf{77}, 066211 (2008).

\bibitem{horvatic} D. Horvatic, H.E. Stanley and B. Podobnik, EPL \textbf{94}, 18007 (2011).

\bibitem{alessio} E. Alessio, A. Carbone, G. Castelli and V. Frappietro, Eur. Phys. J. B \textbf{27}, 197 (2002).

\bibitem{arianos2} S. Arianos, A. Carbone and C. Tuerk, Phys. Rev. E \textbf{84}, 046113 (2011).

\bibitem{carbone4} A. Carbone, in \textit{IEEE TIC-STH 09: IEEE Toronto International Conference on Science and Technology for Humanity} (IEEE, Toronto, CANADA, 2009), p. 691. doi: 10.1109/TIC-STH.2009.5444412

\bibitem{carbone2} A. Carbone and H.E. Stanley, Physica A \textbf{384}, 21 (2007).

\bibitem{carbone3} A. Carbone, G. Castelli and H. Stanley, Phys. Rev. E \textbf{69}, 026105 (2004).

\bibitem{carbone5} A. Carbone and H.E. Stanley, Physica A \textbf{340}, 544 (2004)

\bibitem{shao} Y.H. Shao, G.F. Gu, Z.Q. Jiang and W.X. Zhou, Fractals \textbf{23}, 1550034 (2015).

\bibitem{carbone6} A. Carbone, G. Castelli and H.E. Stanley, \textbf{344}, 267 (2004).

\bibitem{dimatteo} T. Di Matteo, Quant. Financ. \textbf{7}, 21 (2007).

\bibitem{serletis2} A. Serletis and A.A. Rosenberg, Physica A \textbf{380}, 325 (2007).

\bibitem{matsuhita} R. Matsushita, I. Gleria, A. Figueiredo and S. Da Silva, Phys. Lett. A \textbf{368}, 173 (2007).

\bibitem{serletis3} A. Serletis and A.A. Rosenberg, Chaos Soliton. Fract. \textbf{40}, 2007 (2009).

\bibitem{carbone7} A. Carbone, Phys. Rev. E \textbf{76}, 056703 (2007).

\bibitem{carbone8} A. Carbone, B.M. Chiaia, B. Frigo and C. Tuerk, Phys. Rev. E \textbf{82}, 036103 (2010).

\bibitem{turk} C. Tuerk, A. Carbone and B.M. Chiaia, Phys. Rev. E \textbf{81}, 026706 (2010).

\bibitem{kiyono4} Y. Tsujimoto, Y. Miki, S. Shimatani and K. Kiyono, Phys. Rev. E \textbf{93}, 053304 (2016).

\bibitem{abry} P. Abry and D. Veitch, IEEE T. Inform. Theory
  \textbf{44}, 2 (1998).
  
\bibitem{veitch} D. Veitch and P. Abry, IEEE T. Inform. Theory \textbf{45}, 878 (1999).

\bibitem{kiyono5} K. Kiyono, in \textit{2017 International Conference on Noise
    and Fluctuations (ICNF)}, IEEE, Vilnius, LITHUANIA, pp. 1-4 (2017). doi: 10.1109/ICNF.2017.7985951

\bibitem{arianos} S. Arianos and A. Carbone, Physica A 382, 9 (2007).

\bibitem{carbone} A. Carbone and K. Kiyono, Phys. Rev. E \textbf{93}, 063309 (2016).

\bibitem{hoell} M. Hoell and H. Kantz, Eur. Phys. J. B \textbf{88}, 126 (2015).

\bibitem{hoell2} M. Hoell and H. Kantz, Eur. Phys. J. B \textbf{88}, 327 (2015).
   
\bibitem{bardet} J.M. Bardet and I. Kammoun, IEEE T. Inform. Theory \textbf{54}, 2041 (2008).
   
\bibitem{crato} N. Crato, R.R. Linhares and S.R.C. Lopes, J. Stat. Comput. Sim. \textbf{80}, 625 (2010).

\bibitem{heneghan} C. Heneghan and G. McDarby, Phys. Rev. E \textbf{62}, 6103 (2000).

\bibitem{kiyono} K. Kiyono, Phys. Rev. E \textbf{92}, 042925 (2015).

\bibitem{kiyono2} K. Kiyono and Y. Tsujimoto, Phys. Rev. E \textbf{94}, 012111 (2016).

\bibitem{kiyono3} K. Kiyono and Y. Tsujimoto, Physica A: Stat. Mech. Appl. \textbf{462}, 807 (2016)
   
\bibitem{willson} K. Willson, D.P. Francis, R. Wensel, A.J.S. Coats and K.H. Parker, Physiol. Meas. \textbf{23}, 385 (2002).
    
\bibitem{willson2} K. Willson and D.P. Francis, Physiol. Meas. \textbf{24}, N1 (2003).
    
\bibitem{lovstetten} O. L\o vstetten, Phys. Rev. E \textbf{96}, 012141 (2017).
   
\bibitem{hoell3} M. Hoell, H. Kantz and Y. Zhou, Phys. Rev. E \textbf{94}, 042201 (2016).
    
\bibitem{bryce} R. Bryce and K. Sprague, Sci. Rep.-UK \textbf{2}, 315 (2012).

\bibitem{mandelbrot} B.B. Mandelbrot and J.W. van Ness, Siam Rev. \textbf{10}, 422 (1968).

\bibitem{chen3} L. Chen, K.E. Bassler, J.L. McCauley and G.H Gunaratne, Phys. Rev. E \textbf{95}, 042141 (2017).
    
\bibitem{maraun} D. Maraun, H.W. Rust and J. Timmer, Nonlinear Proc. Geoph. \textbf{11}, 495 (2004).
    

    


           

\end{thebibliography}
\end{document}